\documentclass[10pt]{article}
\usepackage[margin=1.25cm]{geometry}
\usepackage{amsmath,amssymb}
\usepackage[utf8x]{inputenc}
\usepackage[table]{xcolor}
\usepackage{siunitx}
\usepackage{graphbox}
\usepackage{authblk}
\usepackage{footmisc}
\usepackage{hyperref}

\newcommand{\pde}[2]{\frac{\partial #1}{\partial #2}}

\newcommand{\de}[2]{\frac{d #1}{d #2}}

\newcommand{\ploseq}[1] {Eq~(\ref{#1})}
\newcommand{\figref}[1] {Fig~\ref{#1}}

\DefineFNsymbolsTM{myfnsymbols}{% def. from footmisc.sty "bringhurst" symbols
  \textasteriskcentered *
  \textdagger    \dagger
}%
\setfnsymbol{myfnsymbols}

\title{Homogenisation for the monodomain model in the presence of microscopic fibrotic structures} 

\author[1,2\footnote{Corresponding Author --- b.lawson@qut.edu.au}]{Brodie A. J. Lawson}
\author[3]{Rodrigo Weber dos Santos}
\author[2]{Ian W. Turner}
\author[4]{Alfonso Bueno-Orovio}
\author[2]{Pamela Burrage}
\author[2,4\footnote{Visiting Professor}]{Kevin Burrage}

\affil[1]{Centre for Data Science, Queensland University of Technology, Queensland, Australia}
\affil[2]{ARC Centre of Excellence for Mathematical and Statistical Frontiers, Queensland University of Technology, Queensland, Australia}
\affil[3]{Graduate Program on Computational Modeling, Universidade de Juiz de Fora, Minas Gerais, Brazil}
\affil[4]{Department of Computer Science, University of Oxford, Oxfordshire, United Kingdom}

\date{}

\begin{document}

\maketitle

\begin{abstract}
Computational models in cardiac electrophysiology are notorious for long runtimes, restricting the numbers of nodes and mesh elements in the numerical discretisations used for their solution. This makes it particularly challenging to incorporate structural heterogeneities on small spatial scales, preventing a full understanding of the critical arrhythmogenic effects of conditions such as cardiac fibrosis. In this work, we explore the technique of homogenisation by volume averaging for the inclusion of non-conductive micro-structures into larger-scale cardiac meshes with minor computational overhead. Importantly, our approach is not restricted to periodic patterns, enabling homogenised models to represent, for example, the intricate patterns of collagen deposition present in different types of fibrosis. We first highlight the importance of appropriate boundary condition choice for the closure problems that define the parameters of homogenised models. Then, we demonstrate the technique's ability to correctly upscale the effects of fibrotic patterns with a spatial resolution of $\SI{10}{\micro\metre}$ into much larger numerical mesh sizes of 100-$\SI{250}{\micro\metre}$. The homogenised models using these coarser meshes correctly predict critical pro-arrhythmic effects of fibrosis, including slowed conduction, source/sink mismatch, and stabilisation of re-entrant activation patterns. As such, this approach to homogenisation represents a significant step towards whole organ simulations that unravel the effects of microscopic cardiac tissue heterogeneities.
\end{abstract}

\section{Introduction}
Computational simulation plays a critical role in our understanding of the functioning of the heart, in particular the complex manifestations of its excitable media dynamics into dangerous arrhythmias \cite{Zhou2018}. An important contributor to many types of arrhythmia is cardiac fibrosis, the pathological formation of scar tissue in the heart~\cite{Nguyen2014}, its arrhythmogenic impacts depending on its spatial organisation on both microscopic~\cite{Hansen2015} and macroscopic~\cite{Zahid2016} scales. However, owing both to limitations of computational feasibility and the resolution of clinical imaging approaches, anatomically-accurate meshes used for the simulation of electrical signalling in the heart typically have spacings of minimum 100 micrometres~\cite{Pathmanathan2012}. This is at least an order of magnitude too large to resolve the complex and varied microscopic structures of fibroblast-deposited collagen that interfere with wave propagation~\cite{deJong2011}. It is therefore vital that such ``sub-mesh scale'' effects of fibrotic obstacles be incorporated into simulations without altering the mesh spacing. Even when not working with an anatomical mesh, this type of upscaling represents a significant computational time saving that may be used alongside other acceleration techniques such as improved numerical techniques (e.g.~\cite{Whiteley2006}) and/or hardware architectures~\cite{Oliveira2018b}.

There has been some progress in incorporating small-scale fibrotic structures into larger-scale cardiac electrophysiology simulations. Through a clever node re-labelling, Costa {\it et al.} were able to incorporate disconnections between neighbouring elements due to strands of collagen~\cite{Costa2014}, although such an approach does not necessarily account for the effects of obstacles {\it within} mesh elements. An alternative approach is based on the mathematical technique of {\it homogenisation}, which explicitly seeks to represent micro-scale effects as modifications to macro-scale problems~\cite{Davit2013}. This has some history in cardiac electrophysiology in the derivation of the well-known bidomain model~\cite{Keener2009, Hand2009, Costa2010, Richardson2011, Grandelius2019}, or its modification in the case of less ordered arrangements of cells~\cite{Kim2010} or to represent the non-ohmic nature of tissue conduction \cite{Hurtado2020}. However, research in using homogenisation to incorporate the effects of fibrosis have been largely limited to spatially periodic structures~\cite{Davidovic2017, Gokhale2018}. Austin {\it et al.}~\cite{Austin2006} used homogenisation based on multigrid techniques to incorporate arbitrarily arranged obstacles into larger-scale simulations, but the analysis of spatiotemporal dynamics was not given. Most importantly, none of these approaches have considered whether or not homogenisation is able to capture the mechanisms through which microscopic obstacles to conduction act as arrhythmia precursors. This is arguably the primary aspect of interest in cardiac electrophysiology simulations.

In this work, we use a volume averaging approach for the incorporation of arbitrary structures of microscopic obstacles into a larger-scale problem. We explore several different choices of boundary conditions for homogenisation sub-problems, in order to determine which is most appropriate for the challenging case of sharp-fronted travelling wave dynamics in the presence of completely non-conductive obstacles of arbitrary shapes. We demonstrate the successful capture of several important pro-arrhythmic effects of cardiac fibrosis by block homogenised models, with one to three orders of magnitude fewer nodes than the corresponding fine-scale models. Indeed, some type of homogenisation is likely necessary for the inclusion of small-scale fibrotic structures into a typical three-dimensional mesh of even a single heart chamber, owing to the number of nodes/elements that would compose the corresponding fine-scale discretisation.

\section*{Materials and Methods}

\subsection*{Simulation of cardiac excitation in obstructed tissue}

The dynamics of cardiac excitation are here governed by the monodomain model \cite{Sundnes2006a}, a simplification of the bidomain model that offers similar quality of predictions in many contexts~\cite{Potse2006, Sundnes2006b, Bourgault2010}. The monodomain model is a parabolic partial differential equation coupled to a set of ordinary differential equations. In the presence of non-conductive obstacles, the monodomain model may be expressed in the form
\begin{equation}
\label{monodomain}
\begin{aligned}
\pde{v}{t} &= \nabla \cdot \Bigl( \mathbf{D} \nabla v \Bigr) - \frac{1}{C_m} ( I_{\mbox{\scriptsize ion}}(v,\mathbf{s}) + I_{\mbox{\scriptsize stim}} ) & & \mbox{within conducting tissue} \\
\de{\mathbf{s}}{t} &= \mathbf{f}(v,\mathbf{s}) & & \mbox{within conducting tissue} \\
0 &= (\mathbf{D} \nabla v) \cdot \hat{\mathbf{n}} & & \mbox{on boundaries (incl. obstructions)}.
\end{aligned}
\end{equation}
Here $v$ is the membrane potential (in lower case to denote a micro-scale variable), $C_m$ is the membrane capacitance and $\mathbf{D}$ is the conductivity tensor. $I_{\mbox{\scriptsize stim}}$ refers to externally supplied stimulus current, and $I_{\mbox{\scriptsize ion}}$ specifies the flow of ions in/out of cardiac cells, which depends on both the membrane potential and a set of state variables $\mathbf{s}$. We choose the reduced version of the ten Tusscher {\it et al.} ionic model~\cite{TenTusscher2006} to define $I_{\mbox{\scriptsize ion}}$ and $f$. This model represents action potentials in human ventricular epicardium using formulations for all of the major Na$^{+}$, K$^{+}$, Ca$^{2+}$ currents involved, but with a set of simplifying assumptions made to greatly reduce computational cost.

\begin{figure}
\centering
\includegraphics[width=12cm]{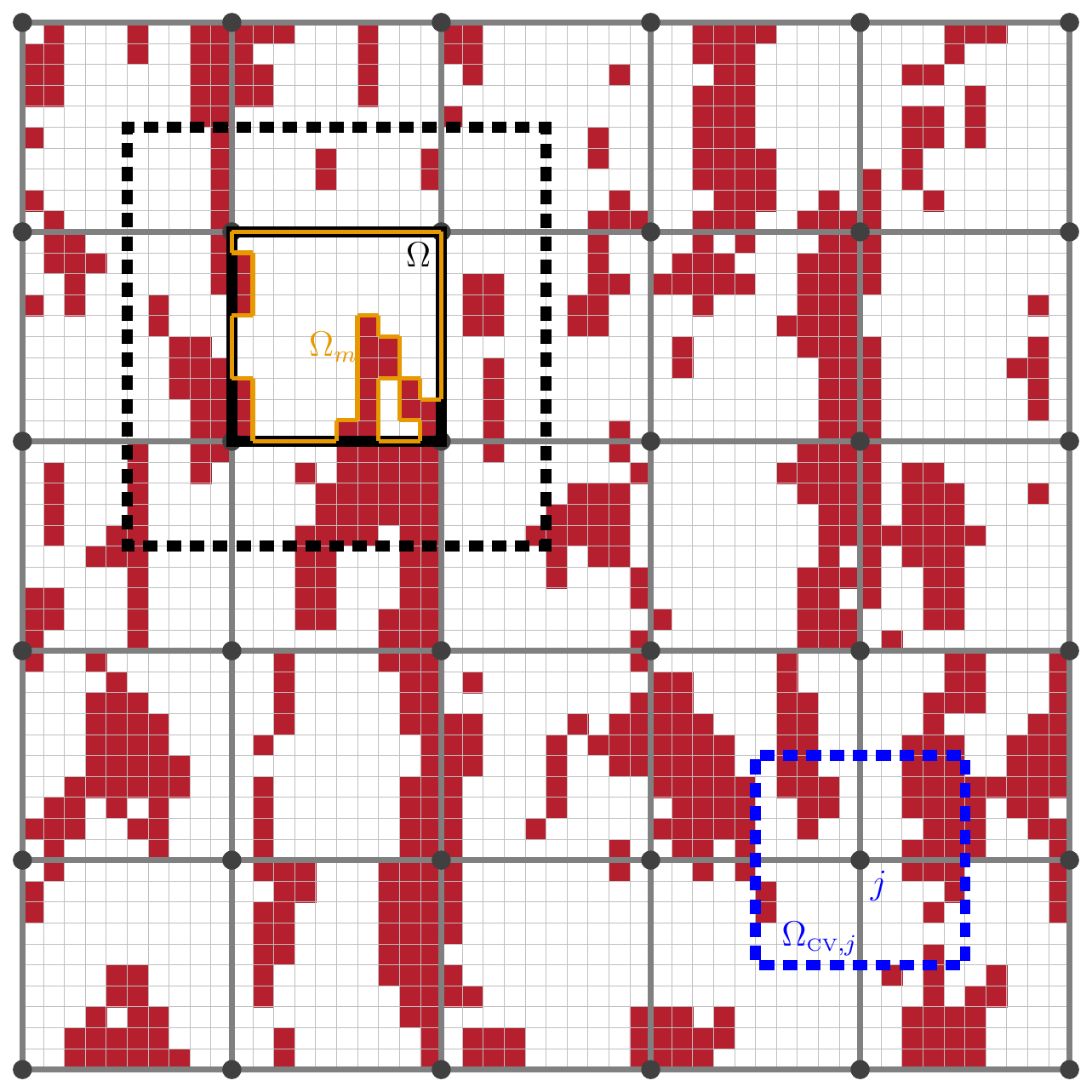}
\caption{The mesh and sub-mesh involved with homogenisation. An example pattern of collagenous obstructions due to fibrosis (dark red) defined on a small-scale mesh, and the larger mesh on which effective conductivity tensors are defined. A single averaging volume {\normalfont ($\Omega$)} is marked in black, along with its associated conductive region, $\Omega_m$, in orange. The black dotted line shows the region on which the closure problem associated with $\Omega$ is solved when a layer of skin is included around the averaging volume (see Methods and Materials). An example control volume used by the finite volume approach used for numerical discretisation is also pictured, in blue.}
\label{fig:mesh_schematic}
\end{figure}

Fibrotic obstructions are defined on a fine-scale grid of spacing $\Delta x = \SI{10}{\micro\metre}$, a similar order to the pixels in histological images indicating the spatial arrangement of collagenous obstacles in cardiac fibrosis~\cite{deJong2011}. The homogenised models we construct seek to represent the effects of these obstacles on a regular grid of a larger scale, as visualised in \figref{fig:mesh_schematic}. We consider homogenised models with $\Delta x = 50, 100, 250, \SI{500}{\micro\metre}$, corresponding to grid spacings common in computational cardiac electrophysiology. Although the current setup uses regular grids with the edges of large scale mesh elements aligned with finescale grid elements, for simplicity, the homogenisation theory presented subsequently does not depend on these choices.

\ploseq{monodomain}, the homogenised equivalent and the closure problems described subsequently, are all solved using a vertex-centred finite volume method. Integrals over control volumes are approximated by exactly integrating linear interpolants, constructed using the nodal values at element vertices. This choice results in a numerical scheme that is similar to that derived using a finite element method without mass lumping, an approach that has been shown to reduce the sensitivity of the monodomain model to the grid spacing used~\cite{Pathmanathan2012}. Timestepping is performed using the second-order generalisation of the Rush--Larsen method presented by Perego and Veneziani~\cite{Perego2009}, with a timestep of $\Delta t = \SI{0.05}{\milli\second}$.

Minimisation of sensitivity of the monodomain model to the spatial discretisation used is paramount for both the evaluation and utilisation of homogenisation, as converting to a homogenised large-scale problem of course incurs a significant change in node spacing. As such, we further correct for the effect of the grid spacing by multiplying all conductivity tensors in our homogenised problems by a constant, such that the conduction velocity in a one-dimensional (non-fibrotic) fibre is consistent with that predicted using the finescale grid spacing ($\SI{10}{\micro\metre}$). We note that this correction factor is selected before, and entirely independent from, the homogenisation process and thus does not act to inflate the perceived performance of the homogenisation itself.

\subsection*{Homogenised monodomain models}

Homogenisation is approached via the method of volume averaging, detailed in \cite{Whitaker1999}. Volume averaging, in the case where non-conductive material is present, makes use of a pair of {\it averaging operators} that average a quantity over the conductive portions, $\Omega_m$, of an averaging volume, $\Omega$. Here, as is typical for this kind of homogenisation, the averaging volume is taken as a single element of the large-scale grid (see \figref{fig:mesh_schematic}). The two averaging operators are
\begin{align*}
& \langle \cdot \rangle = \frac{1}{|\Omega_m|}\int\limits_{\Omega_m} \cdot \, d\Omega_m, & & \mbox{intrinsic average} \\
& \langle \cdot \rangle_{\mbox{\tiny sup}} = \frac{1}{|\Omega|}\int\limits_{\Omega_m} \cdot \, d\Omega_m, & & \mbox{superficial average},
\end{align*}
and are thus linked by the volume fraction of conductive material, \mbox{$\phi = |\Omega_m| / |\Omega|$}, as
\[
\langle \cdot \rangle_{\mbox{\tiny sup}} = \phi \langle \cdot \rangle.
\]
We present a brief, simplified derivation of how these operators are used to formulate homogenised versions of \ploseq{monodomain}.

Consider, as an example, an averaging volume occupied by only a small proportion of cardiac tissue with the remainder occupied by fibrotic obstruction. Even if all of the conductive material on the finescale has a membrane potential $v$ that is above the depolarisation threshold and is about to excite, $\langle v \rangle_{\mbox{\tiny sup}}$ could very well fall below this threshold and hence fails to describe even the ``average'' dynamics of the tissue in the averaging volume. As such, the intrinsic average $\langle v \rangle$ is the quantity we must use in our homogenised model. However, the superficial averaging operator allows us to use the spatial averaging theorem \cite[Ch. 1]{Whitaker1999},
\begin{equation}
    \label{averaging_theorem}
    \biggl\langle \nabla \cdot \mathbf{f} \biggr\rangle_{\mbox{\tiny sup}} = \nabla \cdot \biggl\langle \mathbf{f} \biggr\rangle_{\mbox{\tiny sup}} + \frac{1}{|\Omega|} \int_{\sigma_b} \mathbf{f} \cdot \hat{\mathbf{n}}\, d\sigma_b
\end{equation}
where $\sigma_b$ denotes the surface of the interface between the conductive and non-conducting regions. 

Applying the superficial averaging operator to both sides of the spatially-varying equation in \ploseq{monodomain}, and defining $J = ( I_{\mbox{\scriptsize ion}}(v,\mathbf{s}) + I_{\mbox{\scriptsize stim}} )/C_m$ to simplify notation, then
\begin{equation}
\label{sup_average}
\left\langle \pde{v}{t} \right\rangle_{\mbox{\tiny sup}} = \biggl\langle  \nabla \cdot \Bigl( \mathbf{D} \nabla v \Bigr) \biggr\rangle_{\mbox{\tiny sup}}  - \Bigl\langle J(v, \mathbf{s}) \Bigr\rangle_{\mbox{\tiny sup}}.
\end{equation}
Using \ploseq{averaging_theorem} gives
\[
\left\langle \pde{v}{t} \right\rangle_{\mbox{\tiny sup}} = \nabla \cdot \Bigl\langle \mathbf{D} \nabla v \Bigr\rangle_{\mbox{\scriptsize sup}} + \frac{1}{|\Omega|} \int\limits_{\sigma_b} (\mathbf{D} \nabla v) \cdot \hat{\mathbf{n}} \, d\sigma_b + \Bigl\langle J(v, \mathbf{s}) \Bigr\rangle_{\mbox{\tiny sup}}
\]
however the integral term is seen to be zero due to the boundary condition for the interface between conductive tissue and obstacles. Additionally, in the absence of considering electromechanical coupling we have that obstacles are fixed and so the order of averaging and time differentiation on the left hand side may be reversed. Together, these observations give
\[
\pde{ \langle v \rangle_{\mbox{\tiny sup}} }{t} = \nabla \cdot \biggl( \Bigl\langle \mathbf{D} \nabla v \Bigr\rangle_{\mbox{\tiny sup}} \biggr) + \Bigl\langle J(v, \mathbf{s}) \Bigr\rangle_{\mbox{\tiny sup}}
\]
or in terms of the intrinsic average,
\begin{equation}
\label{nearly_there}
\phi \pde{ \langle v \rangle }{t} = \nabla \cdot \biggl( \phi \Bigl\langle \mathbf{D} \nabla v \Bigr\rangle \biggr) + \phi \Bigl\langle J(v, \mathbf{s}) \Bigr\rangle.
\end{equation}

Finally, we wish to express \ploseq{nearly_there} solely in terms of a macroscopic variable, $V = \langle v \rangle$. In order to achieve this, we define the effective conductivity tensor such that
\begin{equation}
\label{effective_tensor}
\mathbf{D}_{\mbox{\tiny eff}}\nabla V = \Bigl\langle \mathbf{D} \nabla v \Bigr\rangle
\end{equation}
and make the simplifying approximation
\begin{equation}
\label{averaged_source}
\Bigl\langle J(v, \mathbf{s}) \Bigr\rangle \approx J(V, \mathbf{s}).
\end{equation}
This results in the homogenised monodomain model describing the large-scale behaviour of the system,
\begin{equation}
\label{averaged_monodomain}
\begin{aligned}
\phi \pde{V}{t} &= \nabla \cdot \biggl( \phi \mathbf{D}_{\mbox{\tiny eff}} \nabla V \biggr) + \phi J(V, \mathbf{s}) & & \mbox{within tissue}\\
\de{\mathbf{s}}{t} &= \mathbf{f}(V, \mathbf{s}) & & \mbox{within tissue} \\
0 &= (\mathbf{D}_{\mbox{\tiny eff}} \nabla V) \cdot \hat{\mathbf{n}} & & \mbox{on problem boundaries}.
\end{aligned}
\end{equation}
Note that obstacles no longer act through the boundary conditions, but instead through their effect on $\mathbf{D}_{\mbox{\tiny eff}}$ and $\phi$.

We briefly note that a separate technique, the smoothed boundary method, also shifts such boundary conditions into the governing equation to arrive at the formulation \ploseq{averaged_monodomain} \cite{Fenton2005, BuenoOrovio2006}. Volume averaging theory thus serves as a robust means of deriving the smoothed boundary approach. The key difference between the two approaches is their use cases. Smoothed boundary methods use especially fine grids at boundaries in order to accurately resolve their effects, whereas homogenisation by volume averaging instead seeks to represent these effects only on a larger scale, typically to greatly reduce computational demand.

\subsection*{Determination of effective conductivities}

\ploseq{effective_tensor} relates the macroscopic gradient of potential $\nabla V$ to the fine-scale gradient, $\nabla v$. These gradients will change through the course of a simulation of \ploseq{monodomain} or \ploseq{averaged_monodomain}, but fixed conduction tensors can be set by instead solving separate subproblems in which a macroscopic gradient is artificially applied \cite{Wu2002}. Imposing a macroscopic unit gradient in the $i$-th direction ($\nabla V = \mathbf{e}_i$) over the averaging volume, the $i$-th column of the effective conductivity tensor is then given by
\begin{equation}
\label{effective_tensor_columns}
\mathbf{D}_{\mbox{\tiny eff}} \mathbf{e}_i =  \Bigl\langle \mathbf{D} \nabla v_i \Bigr\rangle, \qquad \qquad i = 1,...,d,
\end{equation}
with $\mathbf{e}_i$ the standard basis vectors in the $d$-dimensional space. \ploseq{effective_tensor_columns} applies regardless of the shape of the averaging volume used, or phrased differently, any set of applied gradients may be used to calculate the elements of $\mathbf{D}_{\mbox{\tiny eff}}$ so long as they are in linearly independent directions \cite{Wu2002}.

Each $v_i$ is the solution of a {\it closure} problem, a micro-scale subproblem determining how the imposed macroscopic gradient translates into flow through the material being homogenised (according to the diffusive portion of \ploseq{monodomain}). These problems (and their boundary conditions discussed subsequently) are more naturally expressed in terms of corresponding closure variables,
\begin{equation}
\label{scale_decomposition}
w_i = v_i - x_i,
\end{equation}
the substitution of which into \ploseq{monodomain} results in microscale transport being defined by
\begin{equation}
\label{closure}
\begin{aligned}
0 &= \nabla \cdot \Bigl( \mathbf{D} ( \nabla w_i + \mathbf{e}_i) \Bigr) & & \mbox{within conductive tissue}\\
0 &= \Bigl(\mathbf{D} ( \nabla w_i + \mathbf{e}_i) \Bigr) \cdot \hat{\mathbf{n}} & & \mbox{on boundaries with obstructions}.
\end{aligned}
\end{equation}

Solutions of \ploseq{closure} for each choice of $\mathbf{e}_i$ define the different $w_i$'s, and hence $v_i$'s, with which each column of a large scale mesh element's effective conductivity tensor is then calculated by \ploseq{effective_tensor_columns}. Each element in the large scale problem is assigned its own conductivity tensor via the solution of its own set of independent closure problems, a method referred to as block homogenisation \cite{Durlofsky1992}.

\subsection*{Closure problem boundary conditions}

Arbitrary structures of obstructions in cardiac tissue, such as collagen in cardiac fibrosis, vary significantly in terms of both patterning \cite{deJong2011} and density (for example, within or outside of an infarct region). As such, the assumption of periodicity used in many homogenisation approaches may not be inappropriate. In particular, this assumption will not function when conduction takes place through thin channels stretching across different averaging volumes, as unless those channels happen to align at the opposite ends of an individual volume, the periodic extension implies a non-conducting structure \cite{Davit2013}. This scenario arises in the cardiac electrophysiology context through conductive isthmuses (channels) running through non-conductive scar regions, which are of particular interest as potential substrates for arrhythmia \cite{Asirvatham2014}. 

To address this, we consider alternative boundary conditions for closure problems, \ploseq{closure}, that relax the assumption of periodicity and are commonly used in Laplacian homogenisation \cite{Wu2002, Szymkiewicz2013}. For a rectangular averaging volume with side lengths $L_i$, the set of boundary conditions considered are expressed
\begin{equation}
\label{closure_bcs}
\begin{aligned}
Periodic: & \qquad & w_i( \mathbf{x} + L_j \mathbf{e}_j) &= w_i(\mathbf{x}) & \qquad & \forall j \in {1,\ldots,d} \\
Linear: & \qquad & w_i &= 0 & \qquad & \mathbf{x} \in \partial \Omega \\
Confined: & \qquad & w_i &= 0 & \qquad & \mathbf{x} \in \partial \Omega,\: \mathbf{e}_i \cdot \hat{\mathbf{n}} \neq 0 \\
& \qquad & ( \nabla w_i + \mathbf{e}_i) \cdot \hat{\mathbf{n}} &= 0 & \qquad & \mathbf{x} \in \partial \Omega,\: \mathbf{e}_i \cdot \hat{\mathbf{n}} = 0.
\end{aligned}
\end{equation}
Recalling that the closure problems operate by imposing a macroscopic gradient and considering the resulting flow through the averaging volume, linear boundary conditions may be interpreted as holding all boundaries fixed according to the imposed gradient. Confined conditions, where two opposing boundaries are fixed to maintain the imposed gradient while the remaining boundaries are given no-flux conditions, make Eq (\ref{effective_tensor_columns}-\ref{closure}) a numerical recreation of Darcy's experiments that first derived hydraulic conductivity \cite{SanchezVila1995}.

To further reduce the effects of the assumptions implied by different choices of boundary conditions, we also consider the effects of including a layer of ``skin'' around the averaging volume. This approach extends the domain on which \ploseq{closure} is solved for each averaging volume, beyond the boundaries of that averaging volume \cite{GomezHernandez1990, Wen2000} (see also \figref{fig:mesh_schematic}). As the calculation of effective conductivity through \ploseq{effective_tensor_columns} still only considers the averaging volume, this has the effect of moving the boundaries away from the region used for calculation and thus hopefully minimising boundary condition effects. This is strongly related to the concept of ``oversampling'' in other homogenisation contexts \cite{Henning2013}.

The downside of including skin around averaging volumes is the loss of the guarantee that effective conductivity tensors are symmetric for linear or periodic boundary conditions \cite{Wu2002}. To account for this (as well as asymmetric tensors that may be produced when using confined boundary conditions), we use the algorithm of Higham \cite{Higham1988} to find the symmetric semipositive definite tensor that is closest (in terms of Frobenius norm) to the calculated effective conductivity, $\mathbf{D}_{\mbox{\tiny eff}}$. Briefly, this approach works by calculating the eigendecomposition of the symmetric portion of the initial tensor, $(\mathbf{D}_{\mbox{\tiny eff}} + \mathbf{D}_{\mbox{\tiny eff}}^{T})/2$, zeroing out its negative eigenvalues, and then rebuilding it using this modified eigendecomposition.

\section*{Results}

\subsection*{Non-conductive obstacles complicate boundary condition selection}

The performance of different boundary conditions in homogenisation is often discussed in reviews of the method \cite{SanchezVila1995, Szymkiewicz2013}. However, the case where wholly non-conductive material is present and extends to a significant portion of the boundaries appears to be far less well-considered. We address this by exploring how the three different choices of boundary conditions perform on two particularly illustrative examples, highlighting just how critical the choice of boundary conditions for homogenisation subproblems can be. This then also informs the choice of boundary conditions for formulating homogenised monodomain models in different contexts, as well as the interpretation of the results we observe for the homogenised monodomain models we consider here.

\subsubsection*{Thin Barriers}

\begin{figure}
        \centering
        \includegraphics[height=3.5cm, trim={1.25cm, 7.25cm, 2cm, 6.25cm}, clip]{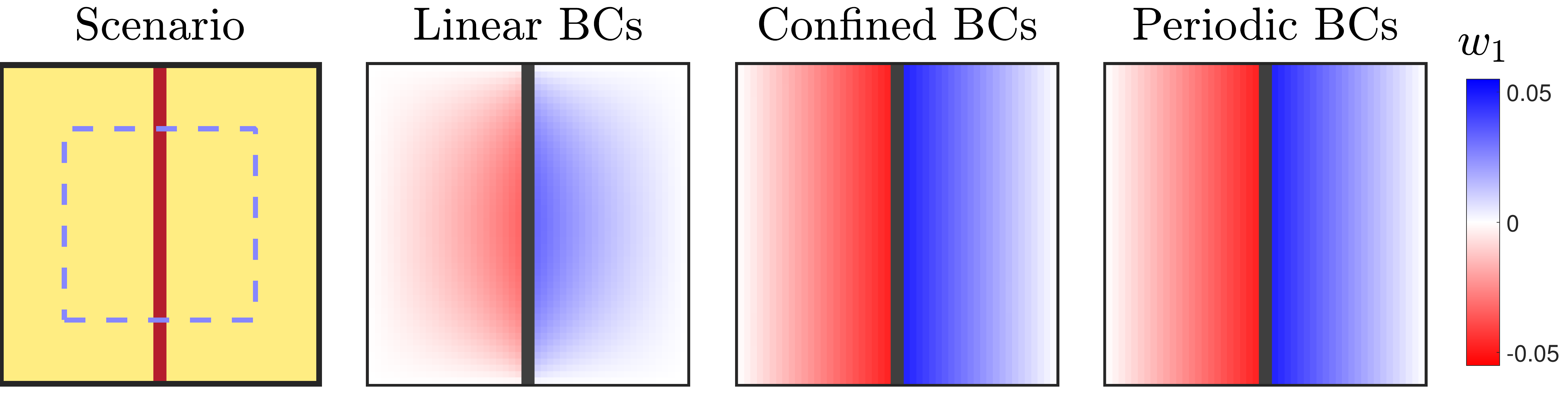}
        
        \vspace{0.5cm}
        
     \begin{tabular}{|c|c|c|c|}
     \hline
     {\bf True} & {\bf Linear BCs} & {\bf Confined BCs} & {\bf Periodic Conditions} \\ \hline
     & & & \\[-0.12cm]
     $\mathbf{D}_{\mbox{\tiny eff}} = \begin{pmatrix} 0 & 0 \\ 0 & 1 \end{pmatrix}$
     & $\mathbf{D}_{\mbox{\tiny eff}} = \begin{pmatrix} 0.486 & 0.000 \\ 0.000 & 1.000 \end{pmatrix}$ & $\mathbf{D}_{\mbox{\tiny eff}} = \begin{pmatrix} 0.000 & 0.000 \\ 0.000 & 1.000 \end{pmatrix}$ & $\mathbf{D}_{\mbox{\tiny eff}} = \begin{pmatrix} 0.000 & 0.000 \\ 0.000 & 1.000 \end{pmatrix}$ \\
     & & & \\[-0.12cm]
     & $\mathbf{D}_{\mbox{\tiny skin}} = \begin{pmatrix} 0.253 & 0.000 \\ 0.000 & 1.000 \end{pmatrix}$ & $\mathbf{D}_{\mbox{\tiny skin}} = \begin{pmatrix} 0.000 & 0.000 \\ 0.000 & 1.000 \end{pmatrix}$ & $\mathbf{D}_{\mbox{\tiny skin}} = \begin{pmatrix} 0.000 & 0.000 \\ 0.000 & 1.000 \end{pmatrix}$ \\
     & & & \\
     \hline
\end{tabular}
    \caption{Performance of different homogenisation boundary conditions (BCs) for a pernicious test case. Conductive tissue is shown in yellow, and fibrotic obstruction in dark red (dark grey in visualisations of the closure problem solution). The dashed blue rectangle indicates the region averaged over when considering the effects of skin. Pictured are the solutions to the closure subproblem \ploseq{closure} for the variable $w_1$, which when averaged together with the $w_2$ solutions (not pictured) result in the conductivity tensors given in the table. Confined and periodic conditions result in a constant gradient solution, such that $\bigl\langle \mathbf{D}(\mathbf{e}_1 + \nabla w_1) \bigr\rangle = \mathbf{0}$ and zero horizontal flow is correctly predicted. Linear boundary conditions result in a degradation of that solution near the boundaries, and this results in an effective tensor that still permits some horizontal flow. Use of skin to shift averaging away from the boundaries helps reduce this effect but falls far short of eliminating it.}
    \label{fig:thin_barrier}
\end{figure}

The first scenario we consider is a single large-scale element composed of isotropically conductive material of constant property, but with a thin strip of obstructive material running vertically along its whole length. This structure is pictured in \figref{fig:thin_barrier}, and it can be intuited that such a structure permits no macroscopic flow in the $x$-direction. In the $y$-direction, although the {\it amount} of macroscopic transport will be lowered slightly by the non-conductive portion, the structure poses no obstacle to vertical conduction. As such, our ``intrinsic'' formulation should define the effective conductivity for this element to be
\[
\mathbf{D}_{\mbox{\tiny eff (true)}} = \begin{pmatrix} 0 & 0 \\ 0 & D \end{pmatrix},
\]
where $D$ is the (scalar) conductivity of the conductive medium. However, not all choices of boundary conditions are able to obtain this simple result.

In \figref{fig:thin_barrier} we present $\mathbf{D}_{\mbox{\tiny eff}}$ as predicted by the different choices of boundary conditions, with and without including a layer of skin. Also presented are the corresponding solutions of \ploseq{closure} when the macroscopic gradient is imposed in the $x$ direction ($w_1$). Confined and periodic conditions produce $w_1$ solutions with a constant gradient from left to right, and this gradient has the appropriate magnitude such that the quantity $\Bigl\langle \mathbf{D}\nabla v_1 \Bigr\rangle = \Bigl\langle \mathbf{D}(\nabla w_1 + \mathbf{e}_1) \Bigr\rangle = \mathbf{0}$. Hence, \ploseq{effective_tensor_columns} results in the correct conductivity tensor, with zero macroscopic flow in the horizontal direction.

On the other hand, when linear boundary conditions fix $w_1 = 0$ along all boundaries, the constant gradient solution is lost (\figref{fig:thin_barrier}). This results in an incorrect effective conductivity tensor that permits considerable flow in the horizontal direction. By avoiding averaging over the boundaries where the solution is most degraded, including skin has a significant positive effect on the conductivity tensor calculated using linear boundary conditions. Even still, a considerable amount of horizontal flow is permitted, and the effect of the vertical barrier on conduction is essentially lost.

\subsubsection*{Diagonally-oriented Channels}

\begin{figure}
\centering
\includegraphics[height=3.5cm, trim={1.2cm, 7.5cm, 24.25cm, 6.5cm}, clip]{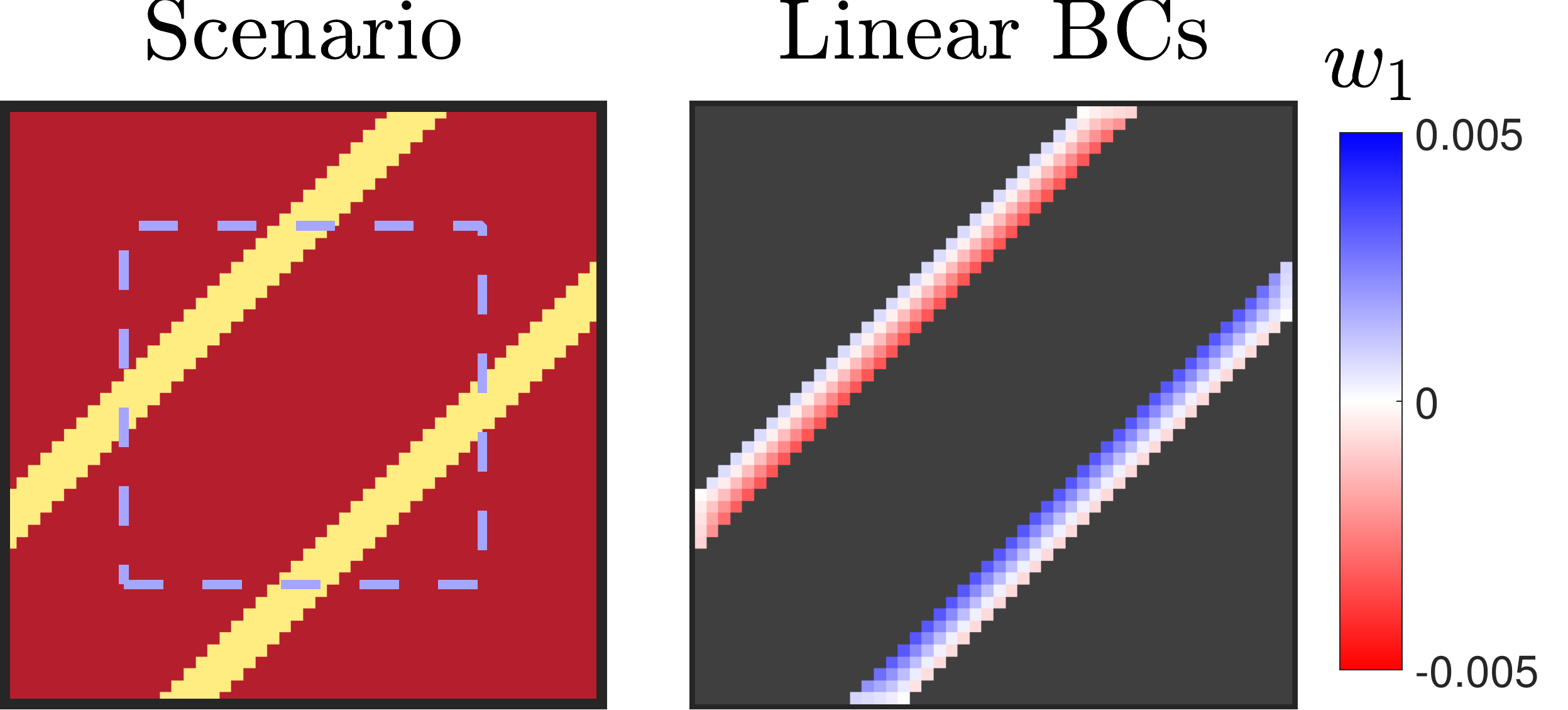}
\includegraphics[height=3.5cm, trim={0cm, 0.3cm, 2.5cm, 0cm}, clip]{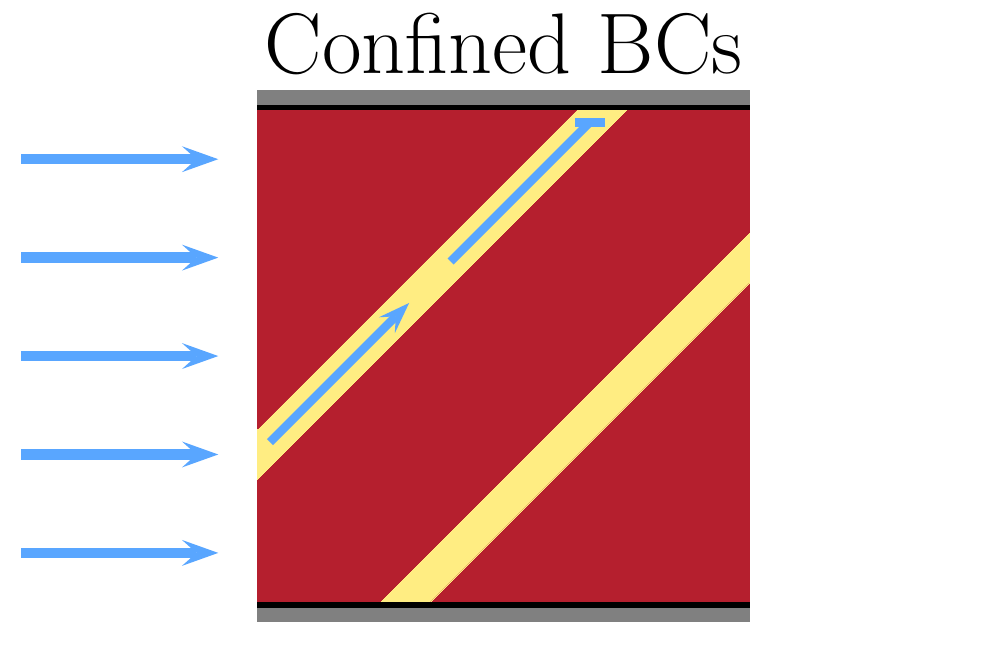}
\includegraphics[height=3.5cm, trim={0cm, 0cm, 0cm, 0cm}, clip]{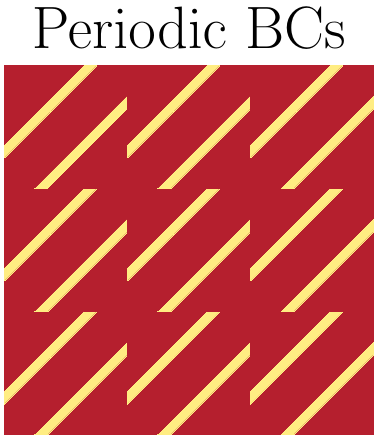}

\vspace{0.5cm}

     \begin{tabular}{|c|c|c|c|}
     \hline
     {\bf True} & {\bf Linear BCs} & {\bf Confined BCs} & {\bf Periodic Conditions} \\ \hline
     & & & \\[-0.12cm]
     $\mathbf{D}_{\mbox{\tiny eff}} = \begin{pmatrix} 0.5 & 0.5 \\ 0.5 & 0.5 \end{pmatrix}$
     & $\mathbf{D}_{\mbox{\tiny eff}} = \begin{pmatrix} 0.518 & 0.481 \\ 0.481 & 0.518 \end{pmatrix}$ & $\mathbf{D}_{\mbox{\tiny eff}} = \begin{pmatrix} 0.000 & 0.000 \\ 0.000 & 0.000 \end{pmatrix}$ & $\mathbf{D}_{\mbox{\tiny eff}} = \begin{pmatrix} 0.000 & 0.000 \\ 0.000 & 0.000 \end{pmatrix}$ \\
     & & & \\[-0.12cm]
     & $\mathbf{D}_{\mbox{\tiny skin}} = \begin{pmatrix} 0.500 & 0.500 \\ 0.500 & 0.500 \end{pmatrix}$ & $\mathbf{D}_{\mbox{\tiny skin}} = \begin{pmatrix} 0.000 & 0.000 \\ 0.000 & 0.000 \end{pmatrix}$ & $\mathbf{D}_{\mbox{\tiny skin}} = \begin{pmatrix} 0.000 & 0.000 \\ 0.000 & 0.000 \end{pmatrix}$ \\
     & & & \\
     \hline
\end{tabular}

    \caption{Performance of different homogenisation boundary conditions (BCs) for a second pernicious test case. Conductive tissue is shown in yellow, and fibrotic obstruction in dark red (dark grey in the visualisation of the closure problem solution). The dashed blue rectangle indicates the region averaged over when considering the effects of skin. Pictured is the solution to the closure problem \ploseq{closure} for the variable $w_1$, and schematic diagrams that indicate the failure of the other types of boundary conditions in this scenario. Again, the evident patterning in the closure problem solution with linear boundary conditions is disrupted at the boundaries. However, disruption is minimal (particularly when skin is included) and the calculated tensor approximates the true tensor. With confined boundary conditions, no flow can pass from left to right when the top and bottom boundaries are blocked (and analogously for when a vertical gradient is imposed), and the calculated tensor is the zero tensor. For periodic conditions, a zero tensor is obtained because the periodic extension of the pattern is seen to be a non-conducting structure.}
    \label{fig:diagonal_channel}
\end{figure}

The second example we consider is a structure for which the scenario is essentially completely reversed. This structure is a pair of diagonally conducting channels through an otherwise non-conductive medium (\figref{fig:diagonal_channel}). Diagonal structures resulting in anisotropy are a popular choice for indicating the effects of the choice of homogenisation boundary conditions on the effective tensors calculated (for example \cite{Renard1997}), but scenarios considered typically treat both materials as conductive. For the situation of conducting channels through a non-conductive medium that we consider here, the effects of boundary condition choice are even more pronounced.

We again treat the conductive material as isotropic and with constant diffusivity $D$. Flow is unimpeded in the direction of the channel, and zero in the direction perpendicular to the channel, and thus by rotational arguments, \[
\mathbf{D}_{\mbox{\tiny eff (true)}} = \begin{pmatrix} \cos \theta & -\sin \theta \\ \sin \theta & \cos \theta \end{pmatrix} \begin{pmatrix} D & 0 \\ 0 & 0 \end{pmatrix}  \begin{pmatrix} \cos \theta & \sin \theta \\ -\sin \theta & \cos \theta \end{pmatrix}  = D \begin{pmatrix} \cos^2 \theta & \cos \theta \sin \theta \\\cos \theta \sin \theta & \sin^2 \theta \end{pmatrix}.
\]
Again, as we take an intrinsic formulation, the width of the channel does not appear in this result. The macroscopic amount of transport is controlled by the volume fraction $\phi$, while $\mathbf{D}_{\mbox{\tiny eff}}$ as we define it describes the {\it character} of this transport.

\figref{fig:diagonal_channel} shows how the different choices of boundary conditions perform in this scenario (with $\theta = \pi/4$ and $D = 1$). The closure problem solution $w_1$ again shows disruption at the boundaries when using linear boundary conditions, but the effect is minor in this case and the effective tensor successfully approximates the true effective conductivity. Using skin further reduces the disruptive effect of the boundary condition, improving the accuracy of the calculated tensor from one decimal place to three.

In contrast, periodic and confined boundary conditions now incorrectly predict zero transport through the averaging volume, with the solution to \ploseq{closure} giving $\Bigl\langle \mathbf{D}(\nabla w_i + \mathbf{e}_i) \Bigr\rangle = \mathbf{0}$. This result is perhaps better understood by considering the physical interpretation of this closure equation, as presented in \figref{fig:diagonal_channel}. Confined boundary conditions imply a Darcy experiment, and thus zero conductivity is predicted as no path leads through the element from left to right or bottom to top. Periodic conditions result in the channel being blocked at both ends, as demonstrated by the periodic extension pictured in the figure. In the case where the diagonal channels are positioned so that their beginnings and ends align in the periodic extension, periodic conditions can then predict the correct tensor (results not pictured).

Together with the results from the previous section regarding thin barriers, we see that when the arrangement of non-conducting obstacles is arbitrary, there exist scenarios for which each type of boundary conditions we have considered produces a poor estimate of macroscopic conductivity. As such, we cannot select a consistently superior choice and instead now consider their performance for practical use of homogenisation in the context of cardiac electrophysiology. Specifically, we explore the potential of homogenised models to capture several key pro-arrhythmic effects of cardiac fibrosis, as represented by the presence of non-conductive fibrotic deposits.

\subsection*{Homogenised monodomain models capture macroscopic excitation propagation in obstructed tissue}

\begin{figure}
\centering
\includegraphics[width=0.99\textwidth, trim={7cm, 0cm, 7cm, 1cm}, clip]{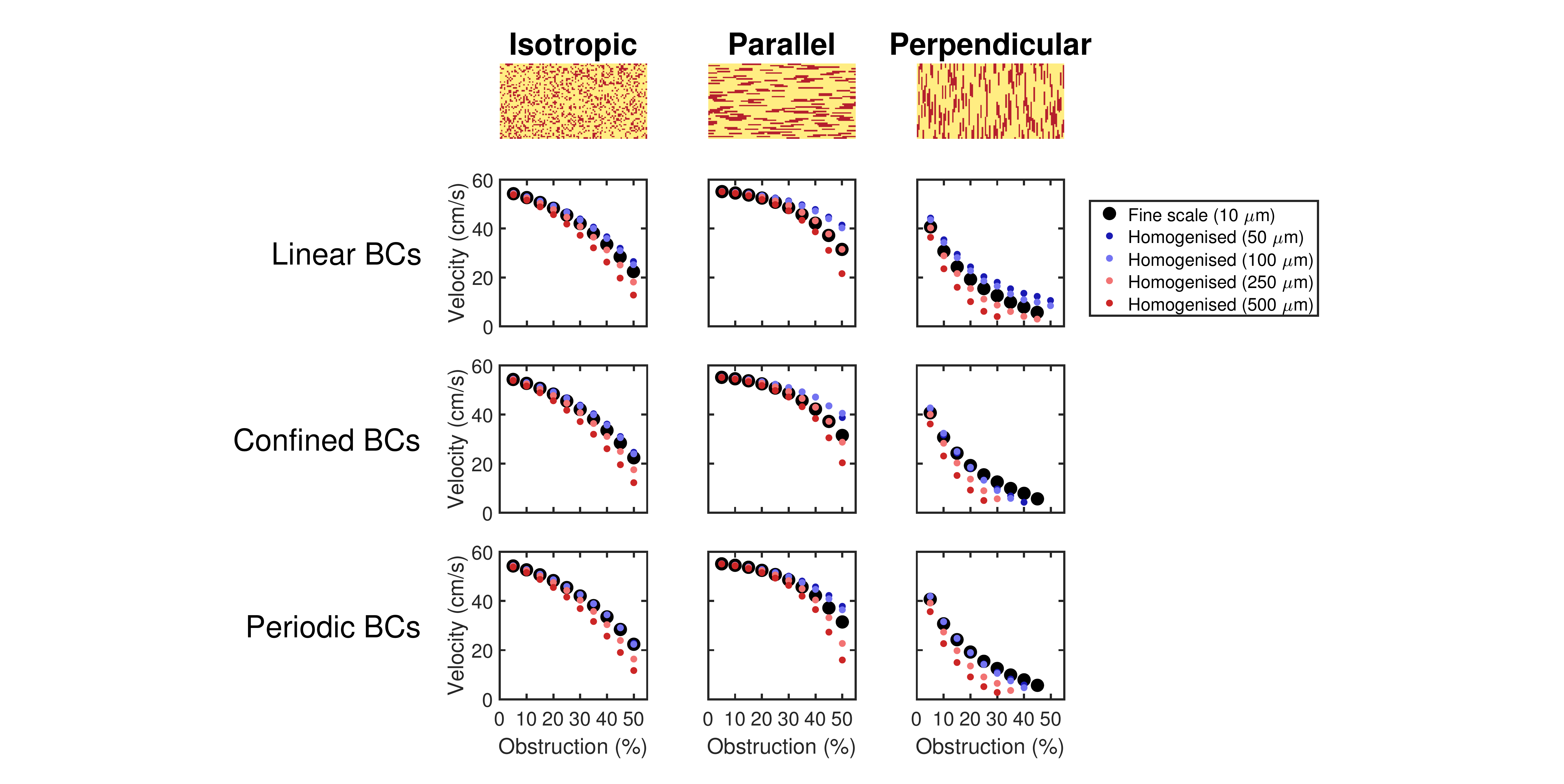}
\caption{ {\bf Prediction by homogenised models of conduction speeds in different types of obstructed tissue.}
\hspace{2.5cm}
{\bf Top Row:} Representative sections of the \SI{5}{\centi\metre} $\times$ \SI{0.5}{\centi\metre} fibres showing the patterns of obstruction considered (obstacles in red). Pictured examples are the case of 25\% obstacles. Waves of excitation move from left to right. 
{\bf Bottom Rows:} Performance of the different types of boundary conditions on the three types of obstruction, for different choices of averaging volume size and boundary conditions for closure subproblems \ploseq{closure}. Where a dot does not appear for a given level of obstruction, this corresponds to a failure to propagate the length of the fibre. Both boundary condition selection and averaging volume size have an important effect on homogenisation performance, with periodic conditions most accurate for smaller averaging volumes $(\Delta x \leq \SI{100}{\micro\metre})$ and linear conditions most accurate for the larger ($\Delta x = \SI{250}{\micro\metre}$). Homogenised models with $\Delta x = \SI{500}{\micro\metre}$ are universally poor for the more challenging, highly obstructed problems. Best overall performance is obtained by using linear boundary conditions and a 25$\!\times\!$25 averaging volume (\SI{10}{\micro\metre} up to \SI{250}{\micro\metre}). Particularly notable is the case of ``perpendicular'' obstructions, where homogenised models using boundary conditions other than linear are prone to over-predicting conduction block. 
}
\label{fig:fibre_results}
\end{figure}

Fibrotic obstructions slow the propagation of cardiac excitation through afflicted tissue, a key component of fibrosis' pro-arrhythmic effect \cite{deBakker1993} as it decreases the ``wavelength'' that governs the survival of dangerous re-entries \cite{Smeets1986}. We therefore use the wavespeed through obstructed tissue as the first test of our homogenisation approach, specifically two-dimensional slices of cardiac tissue measuring \SI{5}{\centi\metre} $\times$ \SI{0.5}{\centi\metre} with an anisotropic conductivity tensor with faster conduction (3:1 ratio for conductivity between the $x$ and $y$ directions) in the direction of propagation to match the faster conduction along cardiac fibres in the heart. In these slices of tissue, we place non-conductive obstructions at random, either \SI{10}{\micro\metre}$\,\!\times\,\!$\SI{10}{\micro\metre} or \SI{90}{\micro\metre}$\,\!\times\,\!$\SI{10}{\micro\metre} in size, with the latter oriented both parallel and perpendicular to the direction of propagation (depicted in \figref{fig:fibre_results}). The effect of these three types of fibrosis on conduction has also been recently considered, separate from the context of homogenisation \cite{Nezlobinsky2020}.

Wavespeeds predicted by the homogenised models match well with fine-scale wavespeeds, but begin to deviate as the amount of fibrotic obstruction increases and paths of conduction become more torturous (\figref{fig:fibre_results}). The performance of different choices of boundary conditions for closure subproblems is comparable, with the superior choice also depending on the size of the averaging volume used. This is a result of the interaction between the error due to the homogenisation, and the numerical consequences of changing the grid spacing (which is exacerbated by the smaller conductivity tensors in highly-obstructed tissue). Error is consistently worst for the largest averaging volumes ($\Delta x = \SI{500}{\micro\metre}$), the case in which homogenisation error is expected to be lowest as the ratio between characteristic length scales grows smaller \cite{Whitaker1999}. As such, this implies that the effect of changing the grid spacing is the predominant source of error.

Perpendicularly aligned fibrosis presents a particularly interesting scenario, as it results in significant reduction in velocity even for small amounts of obstruction, eventually culminating in complete block of conduction when the proportion of obstructive material reaches 50\%. Homogenised models will only be able to predict this block if they feature conductivities small enough to halt conduction, as all path information is lost. The results for this perpendicular fibrosis case mirror those seen in the diagonal channel example we consider in detail (\figref{fig:diagonal_channel}). Specifically, homogenised models attained using linear boundary conditions can potentially over-predict successful conduction, as a connection between any averaging volume boundaries will result in a weakly conductive element (\figref{fig:thin_barrier}) even when the fine-scale structure in fact creates a dead end. On the other hand, confined and periodic boundary conditions result in homogenised models that significantly over-predict conduction block as the homogenisation process can only ``see'' conductive paths that fit with the assumptions that underlie them. Overall, best performance is seen using linear boundary conditions and $\Delta x = \SI{250}{\micro\metre}$, a choice that performs very well across all of the different patternings of fibrosis tested.

\begin{figure}
    \centering
    \includegraphics[width=0.99\textwidth, trim={7cm, 0cm, 7cm, 1cm}, clip]{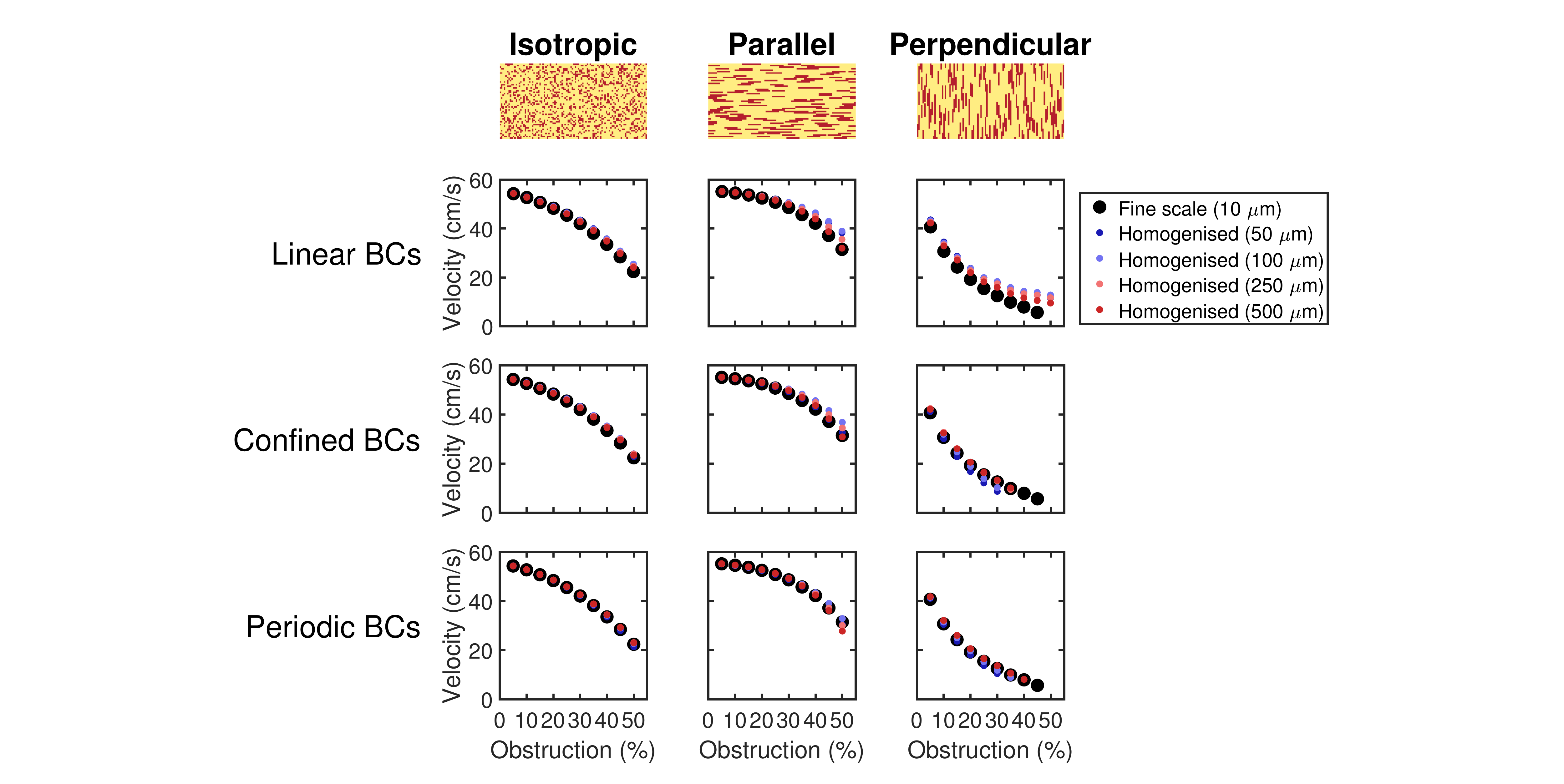}
    \caption{ {\bf Performance of homogenisation for propagation through obstructed tissue when separated from numerical effects.} Displayed conduction velocities are those predicted by homogenised models \ploseq{averaged_monodomain} with different choices of boundary conditions (BCs) and averaging volume size, but solved on the original finescale grid to remove the effects of changing the numerical discretisation. All homogenised models perform well, with best performance obtained using the largest averaging volumes to maximise the difference in length scales between finescale features and the homogenised model. Periodic BCs perform best overall, as the medium is of consistent property and hence admits a representative volume element. When obstacles are aligned perpendicular to the direction of propagation, periodic and confined BCs preemptively predict conduction block for high proportions of obstruction. Linear BCs instead predict conduction even in the case where the finescale simulation blocks.}
    \label{fig:fibre_resultsOG}
\end{figure}

In order to further explore the interaction between homogenisation and numerical error, we also separately consider the results of fine-scale models that have their conductivity fields $\mathbf{D}(\mathbf{x})$ replaced with the effective fields $\mathbf{D}_{\mbox{\tiny eff}}(\mathbf{x})$ and volume fractions $\phi(\mathbf{x})$ obtained through the different block homogenisation approaches. This removes the numerical effects of changing the grid spacing and allows homogenisation error to be more directly examined, but of course does not represent a practical use of homogenisation as there is no computational saving. The results of these tests (\figref{fig:fibre_resultsOG}) present two important conclusions. Firstly, homogenised models perform very well overall, confirming that it is their changed gridsize that produces most of their discrepancy from the equivalent finescale model. Secondly, the overall best-performing boundary conditions now switch to periodic boundary conditions. This is not so surprising, as the tissue fibres here have a consistent patterning of obstacles throughout, and it is therefore reasonable to take a representative volume element and treat the medium as periodic \cite{Whitaker1999}. This shows that in the context of the monodomain model (or other models sensitive to the spatial discretisation used to simulate them), the potentially compensatory balance between homogenisation and grid error must be considered in selecting the parameters of the homogenised model.

\subsection*{Volume averaging allows prediction of source/sink mismatch events}

\begin{figure}[tp]
    \centering
\begin{minipage}[t]{0.03\textwidth}\vspace{1cm}%
    \centering
    {\bf a)}
\end{minipage}
\begin{minipage}[t]{0.48\textwidth}\vspace{0cm}%
    \centering
    \includegraphics[width=0.95\textwidth, trim={1.5cm, 1cm, 0cm, 0cm}, clip]{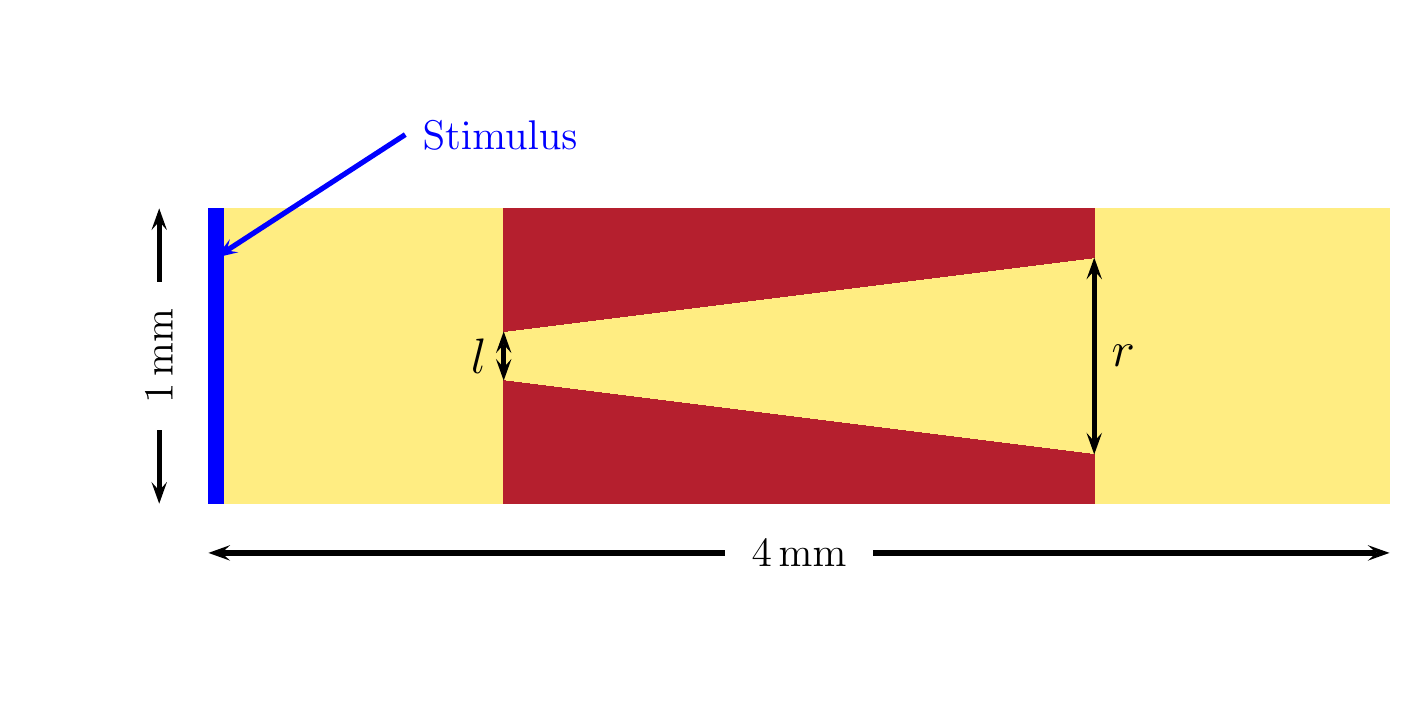}
\end{minipage}
\begin{minipage}[t]{0.03\textwidth}\vspace{1cm}%
    {\bf b)}
\end{minipage}
\begin{minipage}[t]{0.42\textwidth}\vspace{0cm}%
    \centering
    \includegraphics[width=0.95\textwidth, trim={9.75cm, 0cm, 5cm, 2cm}, clip]{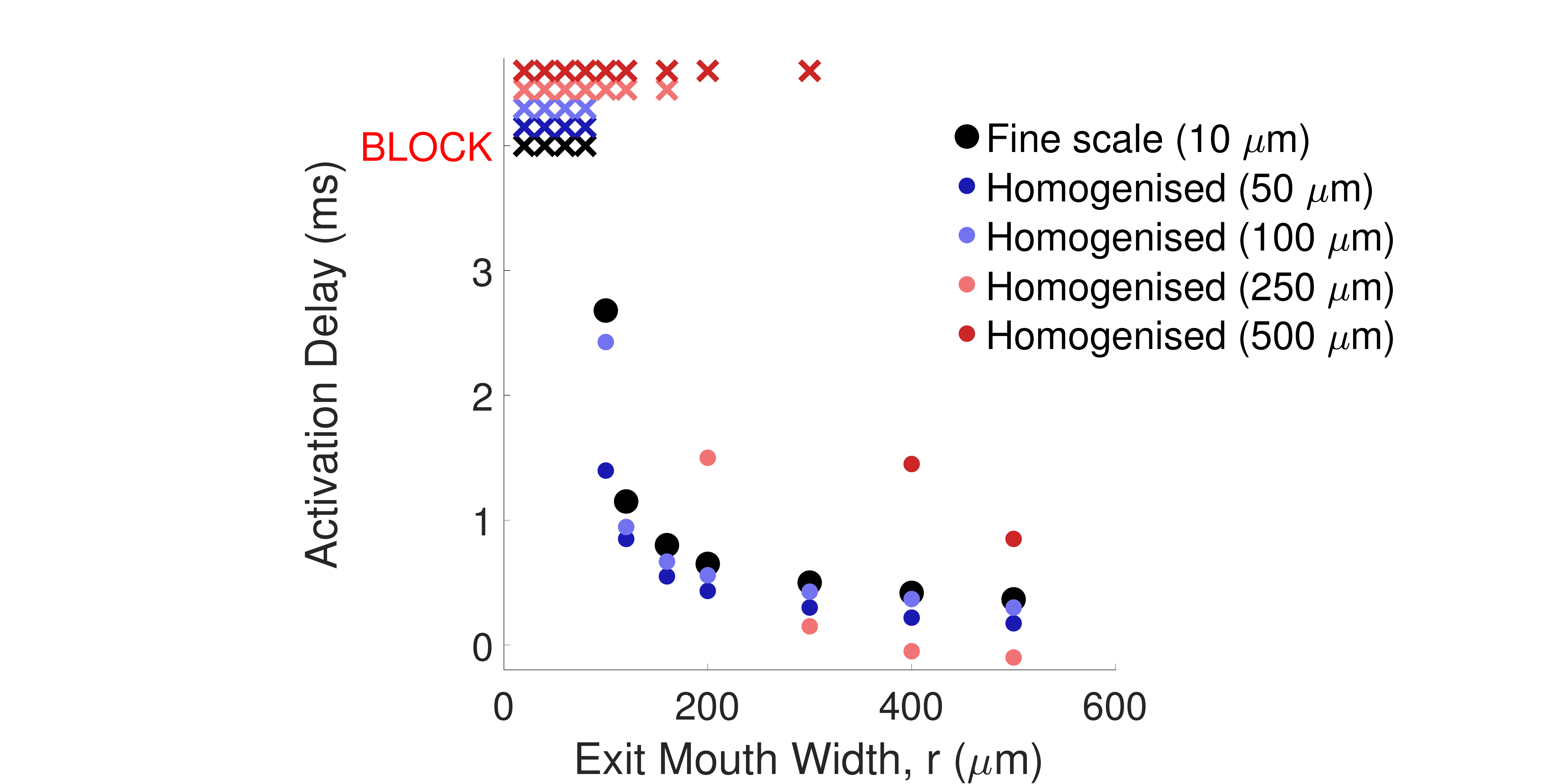}
\end{minipage}
\begin{minipage}[t]{0.03\textwidth}\vspace{2cm}%
    {\bf c)}
\end{minipage}
\begin{minipage}[t]{0.96\textwidth}\vspace{0pt}%
    \centering
    \includegraphics[width=0.95\textwidth, trim={0cm, 0cm, 1cm, 0cm}, clip]{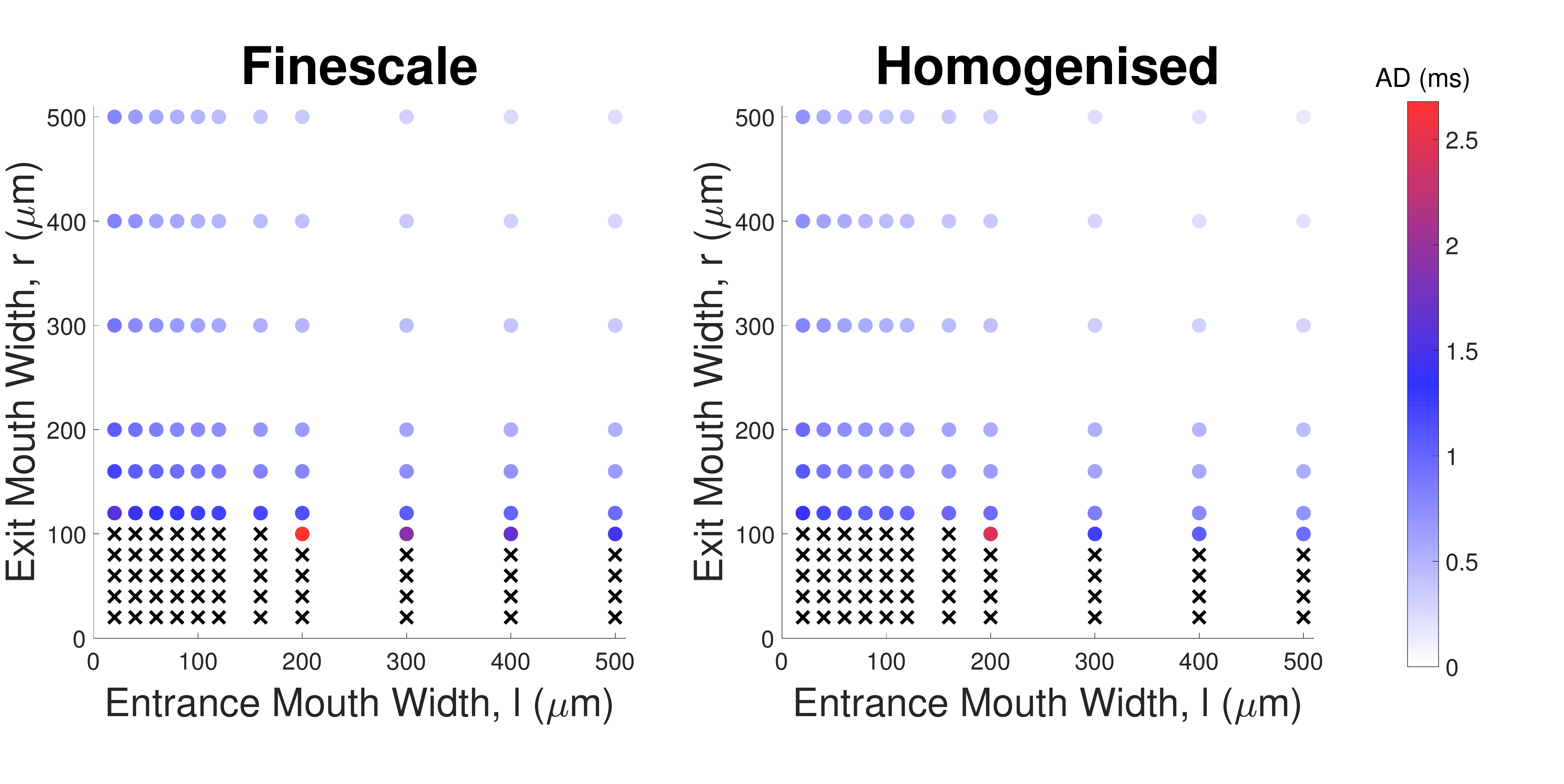}
\end{minipage}
\begin{minipage}{0.03\textwidth}\vspace{1cm}%
    \centering
    {\bf d)}
\end{minipage}
\begin{minipage}{0.96\textwidth}\vspace{0pt}%
    \centering
    \includegraphics[width=14cm, trim={2.5cm, 0cm, 3.75cm, 0cm}, clip]{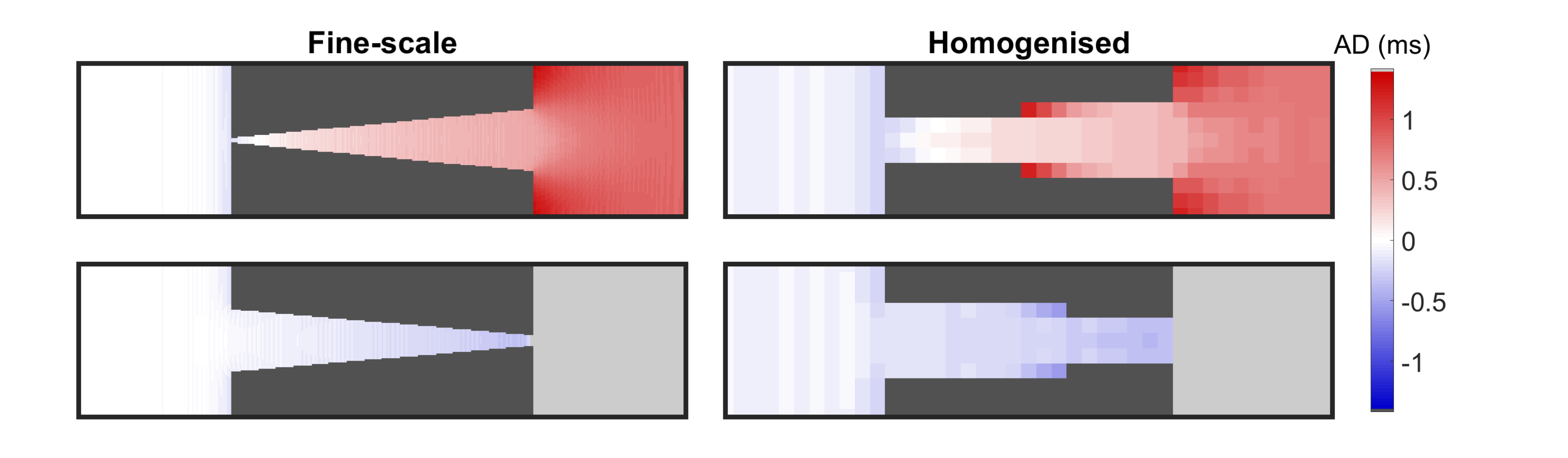}
\end{minipage}
\caption{ {\bf Homogenisation for the capture of source/sink mismatch. }
{\bf a)} The nozzle structure used to evaluate the performance of homogenisation. 
{\bf b)} Activation delay and conduction block as predicted by homogenised models using linear subproblem boundary conditions, for a range of exit widths $r$ ($l = \SI{200}{\micro\metre}$). Performance is good for homogenised models with grid spacing $\Delta x \leq \SI{100}{\micro\metre}$.
{\bf c)} Performance of the homogenised model using linear boundary conditions and $\Delta x = \SI{100}{\micro\metre}$ for multiple combinations of $l$ and $r$. Delay is well estimated by the homogenised model, and block is perfectly predicted in all trialled scenarios.
{\bf d)} Example activation maps for two finescale problems and the equivalent homogenised models (as in {\bf c)}). Patterns of hastened (blue) and delayed (red) activation are well recovered, as is the occurrence of activation failure (light grey).
}
\label{fig:nozzle_results}
\end{figure}

A critical component of the pro-arrhythmic effects of fibrosis is so-called {\it source/sink mismatch}, in which spatial variation in the amount of excitable tissue can create structures that permit conduction in one direction and not another \cite{Nguyen2014}. Unlike the other homogenised models that have been used to represent the impacts of obstacles in cardiac electrophysiology \cite{Davidovic2017, Gokhale2018, Austin2006}, the explicit representation of the local proportion of conductive tissue ($\phi$) in the homogenised models we construct provides them the potential to capture this important electrophysiological dynamic. We examine this potential using a set of small-scale nozzle-like structures (\figref{fig:nozzle_results}a) that produce a delay in activation, or outright block, when activation reaches the end of the structure and attempts to emerge out into unobstructed tissue. Propagation success is seen to depend predominantly on the width of the exit opening, $r$, while the width of the entrance, $l$, has an effect only on borderline cases. Both widths together control the extent of activation delay (calculated by comparing the activation time at the opposite end of the tissue to the activation time when no obstacle is present). We explore how well our homogenised models predict these dynamics, and highlight the fact that the sizes of the averaging volumes trialled are certainly large enough to obscure the fine detail of the structure, and in most cases alter the effective width of the entrance and exit.

Homogenisation performance depends strongly on both the choice of boundary conditions, and the size of the averaging volume. Homogenisation using periodic or confined boundary conditions was found to perform poorly overall (Figures \ref{suppfig:nozzle_results_confined}-\ref{suppfig:nozzle_results_periodic}), most likely due to the poor handling of diagonal transport seen for these choices in our test case (\figref{fig:diagonal_channel}). The nozzle structure considered here also clearly violates the assumption of periodicity. Linear boundary conditions, however, perform well providing the averaging volume is not made too large (\figref{fig:nozzle_results}b). In particular, homogenisation by a factor of ten (resulting in a mesh spacing of \SI{100}{\micro\metre}, consistent with high-fidelity anatomic meshes) proved capable of perfectly predicting the success or failure of propagation for all combinations of $l$ and $r$ values trialled (\figref{fig:nozzle_results}c), and predicted delay accurately in the majority of cases.

Two example maps of the activation delay using this best-performing homogenisation are presented in \figref{fig:nozzle_results}d. Here it is seen that although the precise shape of the nozzle structure is well and truly lost in the homogenised model, the pattern and timing of activation is well recovered. Major discrepancies with the fine-scale problem occur only where a conductive region is spuriously created by the homogenisation process (as any averaging volume containing even a small amount of conductive material will become a non-occupied element in the homogenised model). However, through its choice of effective tensors and incorporation of the volume fraction, the homogenised model is seen to compensate for this effect. In the case of a widening channel, these locations lag in activation, resulting in additional sink to slow propagation. This phenomenologically captures the increasing sink experienced by a wavefront travelling through a widening structure. In the opposite case of a narrowing channel, these regions activate more rapidly and thus replicate the source-favoured balance for a wavefront moving through a narrowing structure.

\subsection*{Spiral wave anchoring can be predicted by homogenised models}

A primary cause of arrhythmia are spiral waves, where cardiac tissue falls into a self-sustaining pattern of continuous re-activation. Depending on both electrophysiological and structural conditions, these spirals may stay fixed rotating about a single region of the tissue or wander about it (with a chance of self-annihilation upon collision with a tissue boundary). Fibrosis can act to stabilise these spiral waves, with even small amounts of diffusely placed obstacles shown to reduce spiral core wander \cite{TenTusscher2007}. Larger non-excitable obstacles can also act as fixed locations to which a spiral wave anchors, and is thus more likely to persist \cite{Davidenko1992}. We explore here whether a homogenised model can still produce this latter important component of fibrosis' pro-arrhythmic effect.

\begin{figure}
    \centering
    \includegraphics[width=0.95\textwidth, trim={6.5cm, 1cm, 1cm, 5.25cm}, clip]{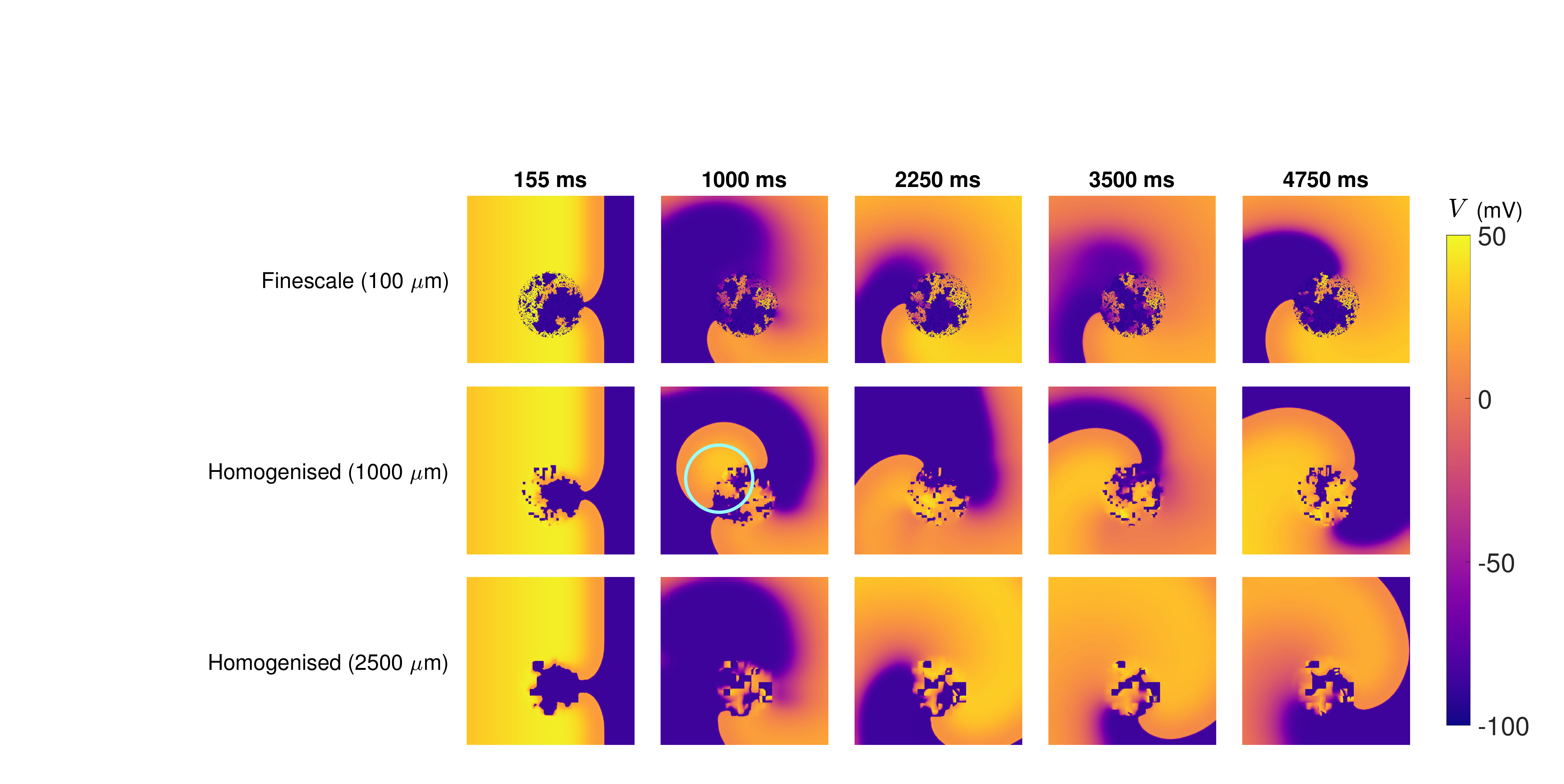}
    \caption{{\bf Anchoring of spiral waves in homogenised models, as shown by simulation snapshots.} A region of 60\% obstruction causes a spiral wave that would otherwise break up to anchor and persist indefinitely. Homogenised models successfully predict this anchoring despite losing the detailed structure of the obstructed region. One homogenised model exhibits transient breakthroughs from the obstructed region not seen in the finescale equivalent (light blue circle). Videos are provided in the supplementary material.}
    \label{fig:spiral_anchoring}
\end{figure}

Spiral waves were simulated in two-dimensional, \SI{6}{\centi\metre} $\!\times\!$ \SI{6}{\centi\metre} slices of tissue, with a ``finescale'' grid spacing of $\Delta x = \SI{100}{\micro\metre}$ used for reasons of computational cost. Spiral waves were initiated using the common cross field stimulus protocol \cite{Beaumont1998}, with a travelling wave initiated at one edge of a two-dimensional slice of tissue, and then a second stimulus triggered in one quadrant of the domain, timed to coincide with the repolarisation front from the first wave. We use the ``steep restitution'' set of parameter values provided by ten Tusscher {\it et al.} \cite{TenTusscher2006}, which cause spiral waves to break up and devolve into irregular patterns of activation. However, then including a region of fibrotic obstruction causes the spiral wave to anchor and be sustained apparently indefinitely without wave breakup (\figref{fig:spiral_anchoring}, first row). 

Following their good performance on the previous scenarios considered, we created homogenised models using linear boundary conditions, and averaging volumes of size 10$\!\times\!$10 and 25$\!\times\!$25 elements. As the base grid in these simulations is already \SI{100}{\micro\metre}, we avoid the numerical effects of significantly larger grid spacings by retaining the base grid but overlaying the tensors obtained by block homogenisation. This approach allows for feasible simulation of sufficient tissue to support spiral waves on the finescale, to which the predictions of homogenised models can be directly compared. We stress, however, that the most practical use case of homogenisation remains incorporating very fine scale ($\sim \!$ \SI{10}{\micro\metre}) structures into typical cardiac meshes ($\sim \!$ 100-\SI{250}{\micro\metre}).

Most immediately, the homogenised models are seen to correctly predict the anchoring of spiral waves to a region of obstruction (\figref{fig:spiral_anchoring}). Although precise positions of wavefronts do not match in individual snapshots, frequencies of rotation measured over multiple rotations are very similar (\SI{3.2}{\hertz} for \SI{100}{\micro\metre} finescale model and \SI{1000}{\micro\metre} homogenised models and \SI{3.1}{\hertz} for \SI{2500}{\micro\metre} homogenised model as estimated from resultant movies, available in the supplement). A noticeable discrepancy between finescale and homogenised models can be seen for the \SI{1000}{\micro\metre} model, where the obstructed region proves too conductive and hence admits breakthroughs that are not seen in the finescale model due to the difference in timing caused by slower traversal through the obstructed region. However, this does not compromise the core behaviours of spiral wave anchoring and frequency of rotation that we want the homogenised models to predict.

\section*{Conclusions}

Homogenisation has seen only limited use in the modelling of fibrosis in cardiac electrophysiology, despite the technique presenting a natural means for incorporating the effects of sub-mesh-scale obstructions such as cardiac fibrosis into existing heart meshes. The dynamic behaviours seen in electrophysiological models, owing to their highly influential and strongly non-linear reaction terms, do present a significant challenge to prototypical homogenisation, which concerns itself only with the calculation of macroscopic transport properties. However, by careful application of the volume averaging theory for non-conductive obstructions \cite{Whitaker1999}, and a thorough consideration of how different choices of closure problem boundary conditions behave in several pernicious cases of interest in this field, we have demonstrated how homogenisation can robustly capture some of the key dynamics promoting arrhythmia.

To demonstrate the method, we have used regular, fine-scale ($\SI{10}{\micro\metre}$) meshes that captured the spatial scale of fibrotic obstructions to conduction, in two dimensions. In practice, a modeller will most likely have a coarser-scale ($100-\SI{250}{\micro\metre}$) mesh, regular or irregular and in two or three dimensions, on which they wish to simulate cardiac activity. In this case, our suggested homogenisation approach would proceed by first temporarily creating a finer-scale grid on which to represent fibrosis, and then solving the set of closure subproblems \ploseq{closure} on this scale (each consisting of a comparatively small number of nodes and elements and able to be processed in parallel). The derived effective conductivities can then be used in the original coarser mesh, along with the volume fractions, through \ploseq{averaged_monodomain}. Although some of the boundary conditions we have considered here (confined and periodic) do not naturally generalise to irregular grids, the fine-scale grid used to represent fibrosis can be freely chosen. As such, a compelling option is to use a regular grid on the finescale, such that boundary conditions and the concept of periodicity can be treated in the same fashion we have demonstrated here. The portions of the finescale mesh in which the irregularly-shaped elements of the coarser mesh are embedded then becomes the ``skin'' that we have demonstrated here to reduce boundary effects.

Overall, homogenisation has been shown here to perform better than might be expected, considering the reaction-dominated dynamics of excitation propagation. Key pro-arrhythmic impacts of fibrotic obstacles to conduction (such as collagenous deposits) have been observed in appropriately homogenised models. This included slowed conduction in fibrosis-afflicted tissue, conduction delay and/or block due to events of source/sink mismatch, and anchoring of spiral waves to heavily obstructed regions. In particular, predictions of conduction slowing were very good for levels of obstruction up to 30-40\%, matching the proportions of collagen that can be seen in typical histological sections of fibrotic tissue \cite{deJong2011}. 

The best-performing homogenised models here used reductions in node count by factors of 100 and 625 compared to simulations explicitly resolving finescale detail, representing significant computational speedup. In three dimensions, this speedup is expected to be even more pronounced. Interpreted differently, the homogenisation approach detailed here makes it feasible to incorporate finescale structures into existing heart meshes already composed of large numbers of nodes and elements, where refinement of the mesh down to the finescale is infeasible due to both computational speed and memory limitations. Although there is a computational cost associated with solving the closure problems that define effective conductivities in a homogenised model, this is a one-off cost, and for the problems considered here this cost was only minor relative to the cost of simulating on the finescale. Where the solution of closure problems threatens to become a bottleneck, they may be solved in parallel or via semi-analytical techniques that further reduce the time required \cite{March2020}.

We have considered the performance of different choices for closure subproblem boundary conditions, both inclusive of, and separate from, the well-known issue of the sensitivity of the monodomain (and bidomain) model to the spatial discretisation \cite{Pathmanathan2012}. Despite their reputation for reduced accuracy \cite{Szymkiewicz2013}, linear boundary conditions proved most accurate and robust overall, here assisted by the inclusion of skin in closure subproblems \cite{GomezHernandez1990} that improved their performance (\figref{fig:diagonal_channel}). Even when periodic boundary conditions are applicable and perform best in terms of homogenisation error, linear boundary conditions become competitive again when the effects of changing gridsize are also incorporated (\figref{fig:fibre_results}). The key weakness of linear boundary conditions in this context is the potential for complete barriers to conduction to ``leak'' (\figref{fig:thin_barrier}), although we consider this superior to the other choices for boundary conditions that have an even higher tendency to over-predict conduction block. Linear boundary conditions were also seen to be clearly superior for the nozzle problem we used to explore source/sink mismatch.

The primary limitation of homogenisation in this context is of course the elimination of the microscopic structure by the homogenisation process. Although we have demonstrated capture of source/sink mismatch and spiral wave anchorage through proof-of-concept experiments, we have not exhaustively tested whether homogenised models can predict the precise manifestations of these effects across the many different types of obstacle arrangement that may be of interest. The further challenge of capturing re-entries that ``live'' on the micro-scale \cite{Hansen2015} has also not been considered here.

In summary, we have demonstrated that when used with care, homogenisation by volume averaging has good potential even for the case where the sensitive and reaction-dominated dynamics of cardiac electrophysiology meet wholly non-conductive material in non-periodic arrangements. Following the result that discrepancies between fine-scale and homogenised models owed mostly to the numerical consequences of the different length scales of the models, this suggests that such homogenisation would perform even better in other settings where the governing equations produce travelling waves with less sharp fronts. The core dynamics of excitation propagation and refractoriness can also be seen, for example, in Ca$^{2+}$ \cite{Lechleiter1991} or cyclic adenosine monophosphate \cite{Dormann2001} signalling in biological systems. The most critical targets for future work are numerical schemes for the monodomain and bidomain equations with further reduced grid sensitivity \cite{Pathmanathan2012, Costa2016}, and/or a new homogenisation paradigm that somehow respects the distinct, tortuous paths through highly obstructed tissue and hence might be capable of generating micro re-entries even in larger-scale homogenised models.

\section*{Acknowledgments}

BAJL, IT, PB and KB are funded by the Australian Research Council Centre of Excellence for Mathematical and Statistical Frontiers (project number CE140100049), an initiative of the Australian government. RWS acknowledges the support by the Brazilian Government via CAPES, CNPq, FAPEMIG, and Universidade Federal de Juiz de Fora, and by the Australian Government via the Endeavour Research Leadership Award from the Department of Education. ABO acknowledges a British Heart Foundation Intermediate Basic Science Fellowship (FS/17/22/32644), and an Impact for Infrastructure Award from the National Centre for the Replacement, Refinement and Reduction of Animals in Research (NC/P001076/1).

\section*{Code Availability}

MATLAB code for the simulation of the monodomain model and the formulation of homogenised models through homogenisation by volume averaging will be made available via GitHub at \texttt{https://github.com/betalawson/fibro-homogenisation}.

\bibliography{homogenisation_fibrosis}

% Temporary counter modification for initial submission version
\setcounter{table}{0}
\renewcommand{\thetable}{S\arabic{table}}%
\setcounter{figure}{0}
\renewcommand{\thefigure}{S\arabic{figure}}%

\clearpage

\section*{Supporting information}

% Include only the SI item label in the paragraph heading. Use the \nameref{label} command to cite SI items in the text.

%%% ADJUST IN FINAL VERSION WHERE ONLY CAPTIONS WILL BE SHOWN

\begin{figure}[!ht]
    \centering
    \begin{minipage}{0.04\textwidth}\vspace{0pt}%
    \centering
    {\bf a)}
    \end{minipage}
    \begin{minipage}{0.94\textwidth}\vspace{0pt}%
    \includegraphics[width=0.65\textwidth, trim={9cm, 0cm, 4cm, 1cm}, clip]{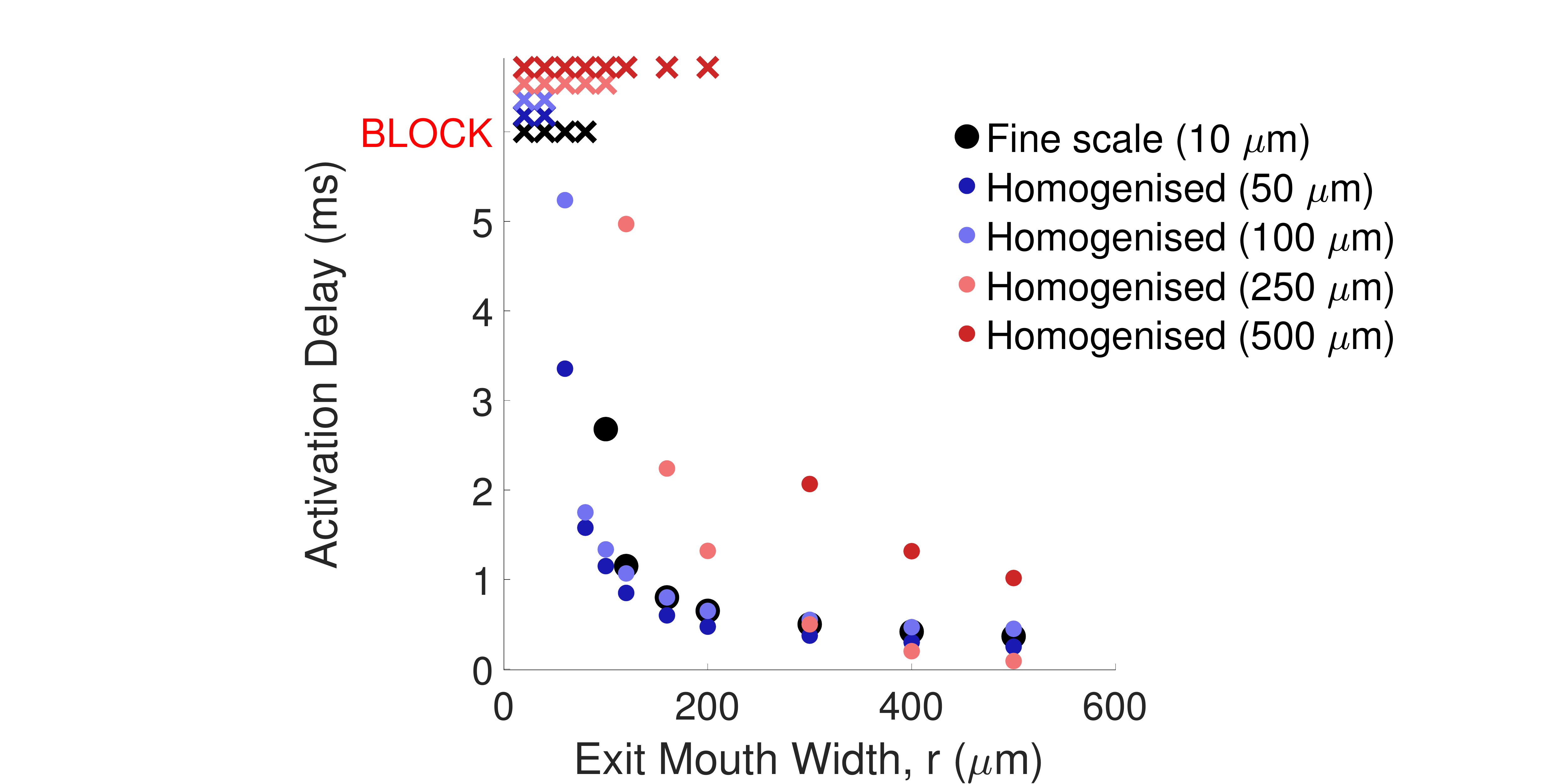}
    \end{minipage}
    \begin{minipage}{0.04\textwidth}\vspace{0pt}%
    \centering
    {\bf b)}
    \end{minipage}
    \begin{minipage}{0.94\textwidth}\vspace{0pt}%
    \centering
    \includegraphics[width=0.95\textwidth, trim={0cm, 2cm, 2.5cm, 0cm}, clip]{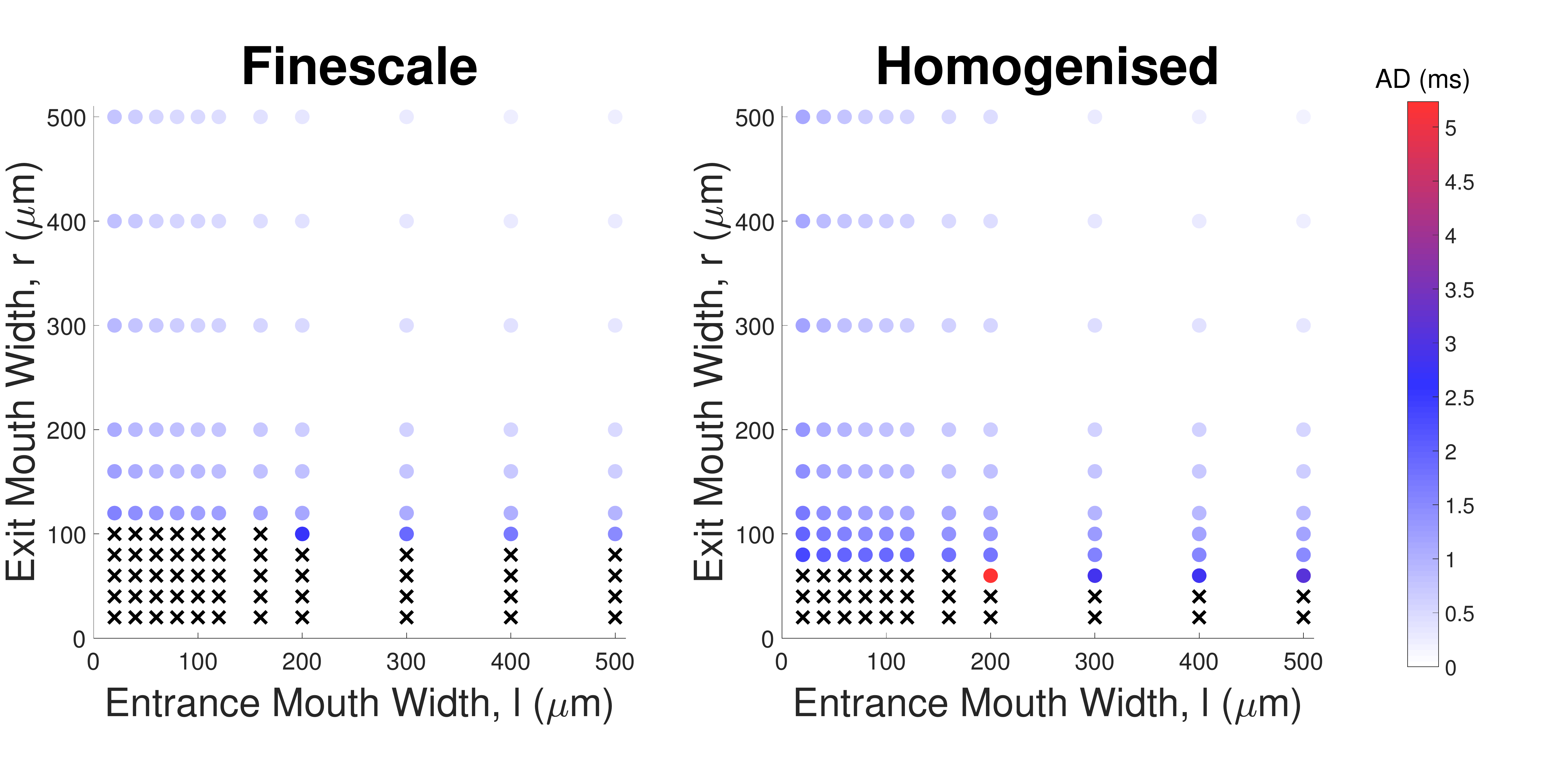}
    \end{minipage}
    \caption{ {\bf The performance of confined boundary conditions for capture of the effects of source/sink mismatch.}
    {\bf a)} Activation delay and conduction block as predicted by homogenised models of  different characteristic lengths, for a range of exit widths $r$ ($l = \SI{200}{\micro\metre}$). Large characteristic lengths for homogenised models result in significant overestimation of delay and block, while the smaller characteristic lengths result in underestimation.
    {\bf b)} Activation delay (AD) and conduction block as predicted by the best-performing homogenised model ($\Delta x = \SI{100}{\micro\metre}$). The rapid transition from low AD to block ($r \sim \SI{100}{\micro\metre}$ in the finescale model) is predicted poorly by even this best-performing homogenised model, indicating a failure of confined boundary conditions.
    }
    \label{suppfig:nozzle_results_confined}
\end{figure}

\clearpage

\begin{figure}[!ht]
    \centering
    \begin{minipage}{0.04\textwidth}\vspace{0pt}%
    \centering
    {\bf a)}
    \end{minipage}
    \begin{minipage}{0.94\textwidth}\vspace{0pt}%
    \includegraphics[width=0.65\textwidth, trim={9cm, 0cm, 4cm, 1cm}, clip]{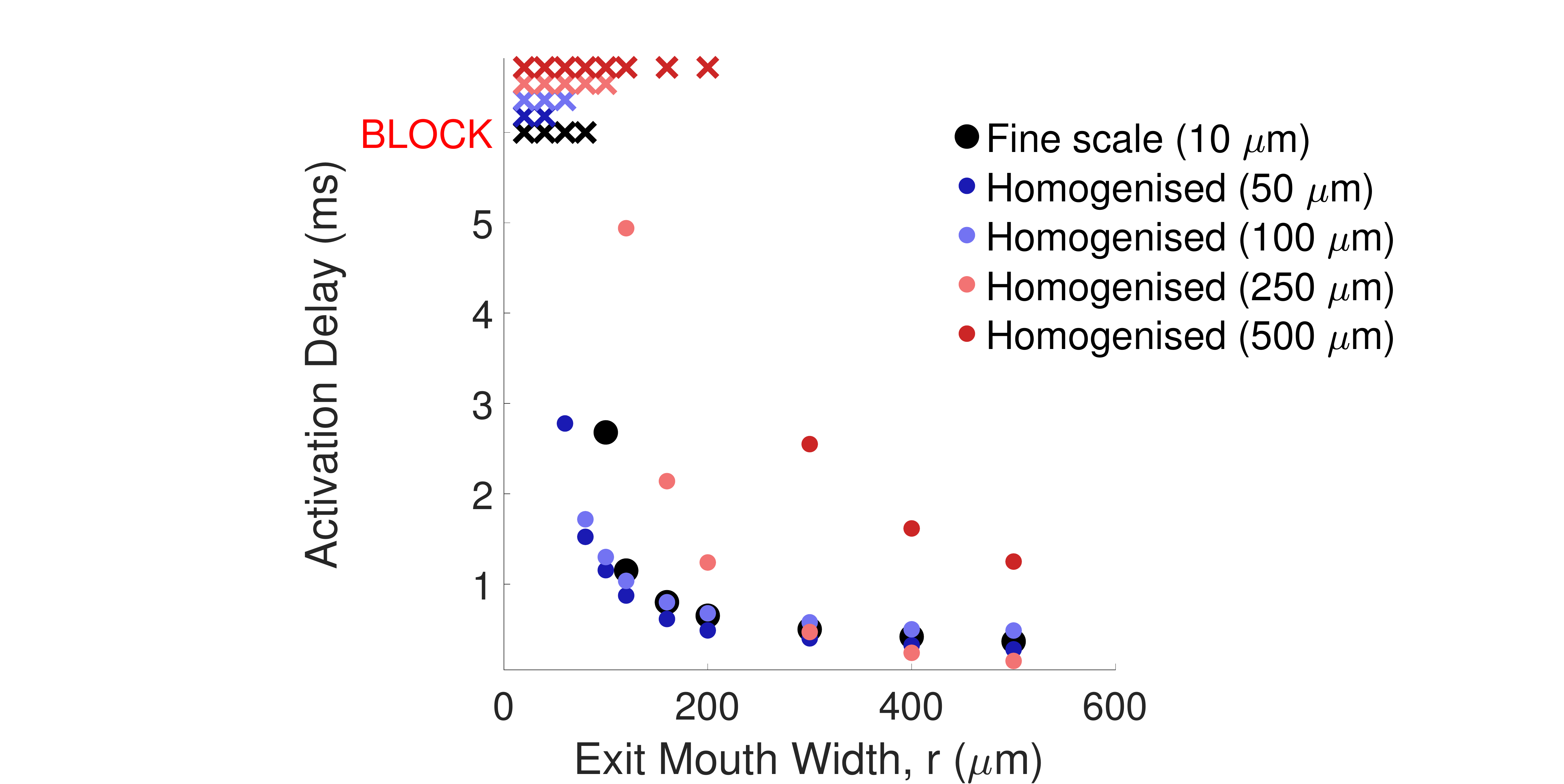}
    \end{minipage}
    \begin{minipage}{0.04\textwidth}\vspace{0pt}%
    \centering
    {\bf b)}
    \end{minipage}
    \begin{minipage}{0.94\textwidth}\vspace{0pt}%
    \centering
    \includegraphics[width=0.95\textwidth, trim={0cm, 2cm, 2.5cm, 0cm}, clip]{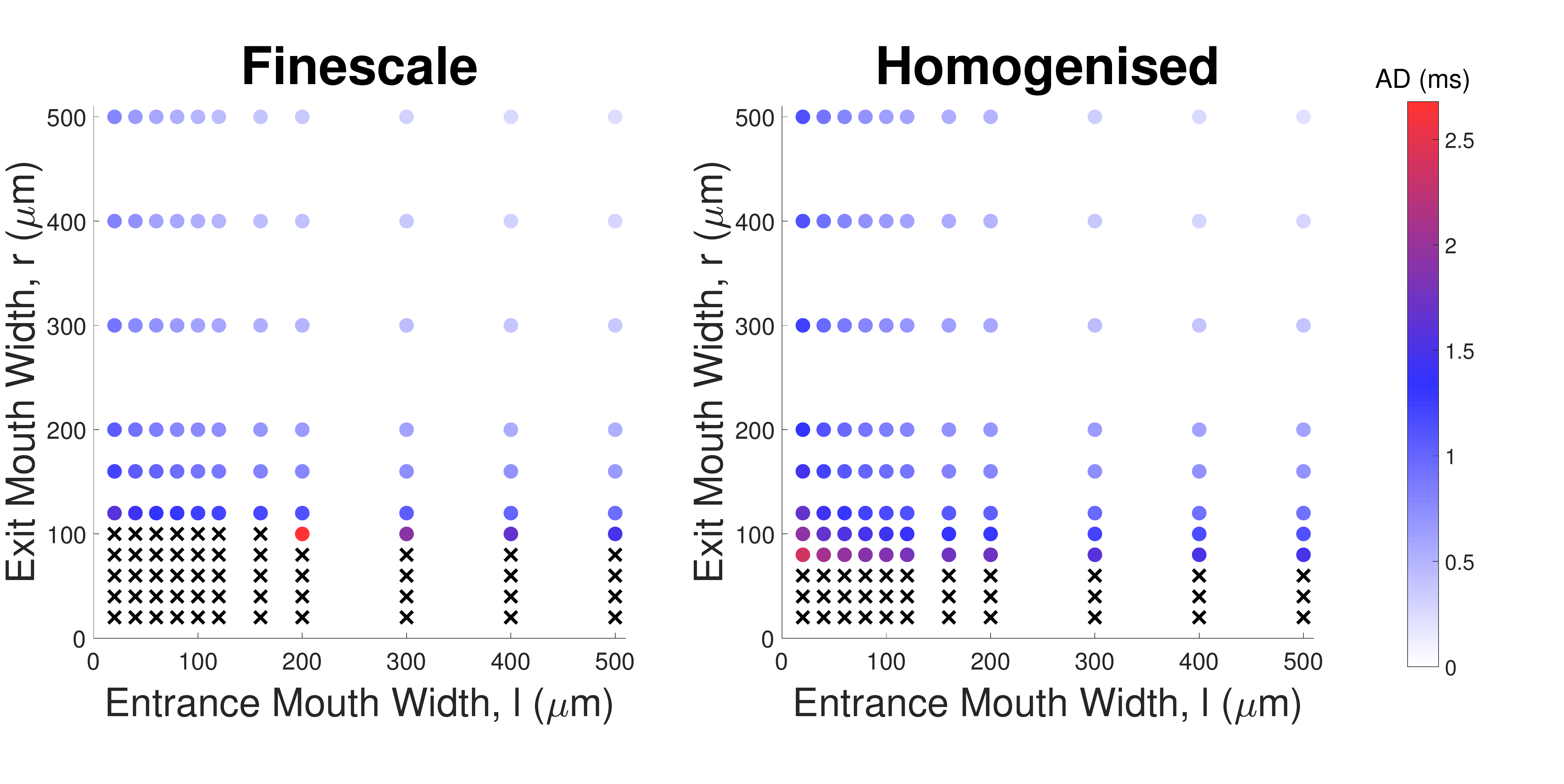}
    \end{minipage}
    \caption{ {\bf The performance of periodic boundary conditions for capture of the effects of source/sink mismatch.}
    {\bf a)} Activation delay and conduction block as predicted by homogenised models of  different characteristic lengths, for a range of exit widths $r$ ($l = \SI{200}{\micro\metre}$). Large characteristic lengths for homogenised models result in significant overestimation of delay and block, while the smaller characteristic lengths result in underestimation.
    {\bf b)} Activation delay (AD) and conduction block as predicted by the best-performing homogenised model ($\Delta x = \SI{100}{\micro\metre}$). The rapid transition from low AD to block ($r \sim \SI{100}{\micro\metre}$ in the finescale model) is predicted poorly by even this best-performing homogenised model, indicating a failure of periodic boundary conditions.}
    \label{suppfig:nozzle_results_periodic}
\end{figure}

\end{document}